\documentclass[12pt,article,onecolumn]{IEEEtran}
\usepackage{amsmath}
\usepackage{amssymb}
\usepackage{cite}
\usepackage{color}
\usepackage{epsfig}
\usepackage{rotating}
\usepackage{mathrsfs}
\usepackage{epsfig}
\usepackage{graphics}
\usepackage{theorem}

\newtheorem{thm}{Theorem}
\newtheorem{definition}{Definition}

\begin{document}

\title{On the Diversity-Multiplexing Tradeoff in Multiple-Relay Network \footnote{Financial supports provided by Nortel, and the corresponding matching
 funds by the Federal government: Natural Sciences and Engineering Research Council of Canada (NSERC)
 and Province of Ontario: Ontario Centres of Excellence (OCE) are gratefully acknowledged.}}

\author{\normalsize
Shahab Oveis Gharan, Alireza Bayesteh, and Amir K. Khandani \\
\small Coding \& Signal Transmission Laboratory\\[-5pt]
\small Department of Electrical \& Computer Engineering\\[-5pt]
\small University of Waterloo \\[-5pt]
\small Waterloo, ON, N2L\ 3G1 \\[-5pt]
\small {shahab, alireza, khandani}@cst.uwaterloo.ca\\}

\date{}
\maketitle \thispagestyle{empty}

\begin{abstract}
This paper studies the setup of a multiple-relay network in which
$K$ half-duplex multiple-antenna relays assist in the transmission
between a/several multiple-antenna transmitter(s) and a
multiple-antenna receiver. Each two nodes are assumed to be either
connected through a quasi-static Rayleigh fading channel, or
disconnected. We propose a new scheme, which we call \textit{random
sequential} (RS), based on the amplify-and-forward relaying. We
prove that for general multiple-antenna multiple-relay networks, the
proposed scheme achieves the maximum diversity gain. Furthermore, we
derive diversity-multiplexing tradeoff (DMT) of the proposed RS
scheme for general single-antenna multiple-relay networks. It is shown that for single-antenna two-hop multiple-access multiple-relay
($K>1$) networks (without direct link between the transmitter(s) and
the receiver), the proposed RS scheme achieves the optimum DMT.
However, for the case of multiple access single relay setup, we show that the RS
scheme reduces to the naive amplify-and-forward relaying and is not
optimum in terms of DMT, while the dynamic decode-and-forward scheme
is shown to be optimum for this scenario \footnote{A Part of this paper,
Theorem 2, is reported in \textit{Library and Archives Canada Technical Report} \cite{relay_tradeoff_1}. Subsequently,
\cite{relay_tradeoff_2} covers Theorems 2 and 3 and
\cite{relay_tradeoff_3} covers Theorems 2, 3, 5 and 6. The materials
of this paper are reported in \cite{relay_tradeoff_4}.}.
\end{abstract}

\section{Introduction}
\subsection {Motivation}
In recent years, relay-assisted transmission has gained significant
attention as a powerful technique to enhance the performance of
wireless networks, combat the fading effect, extend the coverage,
and reduce the amount of interference due to frequency reuse. The
main idea is to deploy some extra nodes in the network to facilitate
the communication between the end terminals. In this manner, these
supplementary nodes act as spatially distributed antennas for the
end terminals. More recently, cooperative diversity techniques have
been proposed as candidates to exploit the spatial diversity offered
by the relay networks (for example, see \cite{laneman, azarian,
yuksel, khisti}). A fundamental measure to evaluate the performance
of the existing cooperative diversity schemes is the
diversity-multiplexing tradeoff (DMT) which was first introduced by
Zheng and Tse in the context of point-to-point MIMO fading channels
\cite{zheng_tse}. Roughly speaking, the diversity-multiplexing
tradeoff identifies the optimal compromise between the ``transmission
reliability" and the``data rate"  in the high-SNR regime.

In spite of all the interest in relay networks, none of the existing
cooperative diversity schemes is proved to achieve the optimum DMT.
The problem has been open even for the simple case of half-duplex
single-relay single-source single-destination single-antenna setup.
Indeed, the only existing DMT achieving scheme for the single-relay
channel reported in \cite{yuksel} requires knowledge of CSI (channel
state information) for all the channels at the relay node.

\subsection{Related Works}

The DMT of relay networks was first studied by Laneman {\em et al.}
in \cite{laneman} for half-duplex relays. In this work, the authors
prove that the DMT of a network with single-antenna nodes, composed
of a single source and a single destination assisted with $K$
half-duplex relays, is upper-bounded by\footnote{Throughout the
paper, for any real value $a$, $a^+\equiv\max\left\{0, a \right\}$.}
\begin{equation}
d(r) = (K+1) (1-r)^{+}. \label{eq:ub}
\end{equation}
This result can be established by applying either the
multiple-access or the broadcast cut-set bound \cite{cover_book} on
the achievable rate of the system. In spite of its simplicity, this
is still the tightest upper-bound on the DMT of the relay networks.
The authors in \cite{laneman} also suggest two protocols based on
decode-and-forward (DF) and amplify-and-forward (AF) strategies for
a single-relay system with single-antenna nodes. In both protocols,
the relay listens to the source during the first half of the frame,
and transmits during the second half. To improve the spectral
efficiency, the authors propose an incremental relaying protocol in
which the receiver sends a single bit feedback to the transmitter
and to the relay to clarify if it has decoded the transmitter's
message or needs help from the relay for this purpose. However, none
of the proposed schemes are able to achieve the DMT upper-bound.

The non-orthogonal amplify-and-forward (NAF) scheme, first proposed
by Nabar {\em et al.} in \cite{nabar_bolcskei}, has been further
studied by Azarian {\em at al.} in \cite{azarian}. In addition to
analyzing the DMT of the NAF scheme, reference \cite{azarian} shows
that NAF is the best in the class of AF strategies for
single-antenna single-relay systems. The dynamic decode-and-forward
(DDF) scheme has been proposed independently in
\cite{azarian,mitran_tarokh,katz_shamai} based on the DF strategy.
In DDF, the relay node listens to the sender until it can decode the
message, and then re-encodes and forwards it to the receiver in the remaining time.
Reference \cite{azarian} analyzes the DMT of the DDF scheme and
shows that it is optimal for low rates in the sense that it achieves
(\ref{eq:ub}) for the multiplexing gains satisfying $r \leq 0.5$.
However, for higher rates, the relay should listen to the
transmitter for most of the time, reducing the spectral efficiency.
Hence, the scheme is unable to follow the upper-bound for high
multiplexing gains. More importantly, the generalizations of NAF and
DDF for multiple-relay systems fall far from the upper-bound,
especially for high multiplexing gains.

Yuksel {\em et al.} in \cite{yuksel} apply compress-and-forward (CF)
strategy and show that CF achieves the DMT upper-bound for multiple-antenna
half-duplex single-relay systems. However, in their proposed scheme,
the relay node needs to know the CSI of all the channels in the
network which may not be practical.

Most recently, Yang {\em et al.} in \cite{yang_belfiore2} propose a
class of AF relaying scheme called slotted amplify-and-forward (SAF)
for the case of half-duplex multiple-relay ($K>1$) and single
source/destination setup. In SAF, the transmission frame is divided
into $M$ equal length slots. In each slot, each relay transmits a
linear combination of the previous slots. Reference
\cite{yang_belfiore2} presents an upper-bound on the DMT of SAF and
shows that it is impossible to achieve the MISO upper-bound for
finite values of $M$, even with the assumption of full-duplex
relaying. However, as $M$ goes to infinity, the upper-bound meets
the MISO upper-bound. Motivated by this upper-bound, the authors in
\cite{yang_belfiore2} propose a half-duplex sequential SAF scheme.
In the sequential SAF scheme, following the first slot, in each
subsequent slot, one and only one of the relays is permitted to
transmit an amplified version of the signal it has received in the
previous slot. By doing this, the different parts of the signal are
transmitted through different paths by different relays, resulting
in some form of spatial diversity. However, \cite{yang_belfiore2}
could only show that the sequential SAF achieves the MISO
upper-bound for the setup of non-interfering relays, i.e. when the
consecutive relays (ordered by transmission times) do not cause any
interference on one another.

Apart from investigating the optimum diversity-multiplexing tradeoff  for relay networks, recently, other aspects of the relay
networks has also been studied (for example, see \cite{kramer,
xie, madsen, gastpar2, avesti_outage, avesti_wireless_deterministic,
gupta_kumar, xie_kumar, gastpar, nabar, shahab_parallel, nabar3,
bolcskei}). \cite{kramer, xie} develop new coding schemes
based on Decode-and-Forward and Compress-and-Forward relaying
strategies for relay networks. Avestimehr {\em et al.} in
\cite{avesti_outage} study the outage capacity of the relay channel for
low-SNR regime and show that in this regime, the bursty
Amplify-and-Forward relaying protocol achieves the optimum outage.
Avestimehr {\em et al.} in \cite{avesti_wireless_deterministic}
present a linear deterministic model for the wireless relay network
and characterize its exact capacity. Applying the capacity-achieving
scheme of the corresponding deterministic model, the authors in
\cite{avesti_wireless_deterministic} show that the capacity of
wireless single-relay channel and the diamond relay channel can be
characterized within 1 bit and 2 bits, respectively, regardless of
the values of the channel gains. The scaling law capacity of large
wireless networks is addressed in \cite{gupta_kumar, xie_kumar,
gastpar, nabar, shahab_parallel, nabar3, bolcskei}. Gastpar {\em et
al.} in \cite{gastpar} prove that employing AF relaying achieves the
capacity of the Gaussian parallel single-antenna relay network for
asymptotically large number of relays. Bolcskei {\em et al.} in
\cite{nabar} extend the work of \cite{gastpar} to the parallel
multiple-antenna relay network and characterize the capacity of
network within $O(1)$, for large number of relays. Oveis Gharan {\em
et al.} in \cite{shahab_parallel} propose a new AF relaying scheme
for parallel multiple-antenna fading relay networks. Applying the
proposed AF scheme, the authors in \cite{shahab_parallel}
characterize the capacity of parallel multiple-antenna relay
networks for the scenario where either the number of relays is large
or the power of each relay tends to infinity.

Recently, in a parallel and independent work by Kumar \textit{et al}\cite{vkumar}\footnote{After the completion of this work, the authors became aware of \cite{vkumar}.} the possibility of achieving the optimum DMT is shown in single-antenna half-duplex relay networks with some graph topologies including KPP , KPP(I), KPP(D) graphs for $K \geq 3$. A KPP graph is a directed graph consisted of $K$ vertex-disjoint paths each with the length greater than one, connecting the transmitter to the receiver. KPP(I) is a directed graph consisted of $K$ vertex-disjoint paths each with length greater than one, connecting the transmitter to the receiver, and possible edges between different paths. KPP(D) is a directed graph consisted of $K$ vertex-disjoint paths each with length greater than one, and a direct path connecting the transmitter to the receiver. It is worth mentioning that in all the mentioned graph topologies, the upper-bound of DMT is achieved by a cut-set of the MISO or SIMO form, i.e. all edges crossing the cut are originated from or destined to the same vertex. Also, they show that the maximum diversity can be achieved in a general multiple-antenna multiple relays network.

\subsection{Contributions}

In this paper, we propose a new scheme, which we call random
sequential (RS), based on the SAF relaying for general
multiple-antenna multi-hop networks. The key elements of the
proposed scheme are: 1) signal transmission through sequential paths
in the network, 2) path timing such that no non-causal interference
is caused from the transmitter of the future paths on the receiver
of the current path, 3) multiplication by a random unitary matrix at
each relay node, and 4) no signal boosting in amplify-and-forward
relaying at the relay nodes, i.e. the received signal is amplified by a coefficient with the absolute value of at most 1. 
Furthermore, each relay node knows the CSI of its corresponding
backward channel, and the receiver knows the equivalent end-to-end
channel. We prove that this scheme achieves the maximum diversity
gain in a general multiple-antenna multiple-relay network (no
restriction imposed on the set of interfering node pairs).
Furthermore, we derive the DMT of the RS scheme for general
single-antenna multiple-relay networks. Specifically, we derive: 1)
the exact DMT of the RS scheme under the condition of
``non-interfering relaying'', and 2) a lower-bound on the DMT of the
RS scheme (no conditions imposed). Finally, we prove that for
single-antenna multiple-access multiple-relay networks (with $K>1$
relays) when there is no direct link between the transmitters and
the receiver and all the relays are connected to the transmitter and
to the receiver, the RS scheme achieves the optimum DMT. However,
for two-hop multiple-access single-relay networks, we show that the proposed
scheme is unable to achieve the optimum DMT, while the DDF scheme is shown to perform
optimum in this scenario.

It is worth mentioning that the optimality results in this paper can easily be applied to the case of KPP and KPP(D) graphs introduced in \cite{vkumar}. However, the proof approach we use in this paper is entirely different from that of used in \cite{vkumar}; Our proofs are based on the matrix inequalities while the proofs of \cite{vkumar} are based on information-theoretic inequalities. Furthermore, \cite{vkumar} shows the achievability of the maximum diversity gain in a general multiple-antenna multiple-relay network by considering a multiple-antenna node as multiple single-antenna nodes and using just one antenna at each time, while in our proof we show that the proposed RS scheme in general can achieve the maximum diversity also in the MIMO form and by using all the antennas simultaneously. 
Finally, the achievability of the linear DMT between the points $(0,d_{\max})$ and $(1,0)$ in  single-antenna layered network and directed acyclic graph network with full-duplex relays  is independently shown as a remark of Theorems 1 and 4 in our paper, respectively.

The rest of the paper is organized as follows. In section II, the
system model is introduced. In section III, the proposed random
sequential scheme (RS) is described. Section IV is dedicated to the
DMT analysis of the proposed RS scheme. Section V proves the
optimality of the RS scheme in terms of diversity gain in general
multiple-antenna multiple-relay networks. Finally, section VI
concludes the paper.

\subsection{Notations}
Throughout the paper, the superscripts $^T$ and $^H$  stand for
matrix operations of transposition and conjugate transposition, respectively. Capital bold letters
represent matrices, while lowercase bold letters and regular letters
represent vectors and scalars, respectively. $\|\mathbf{v}\|$
denotes the norm of vector $\mathbf{v}$ while $\|\mathbf{A}\|$
represents the Frobenius norm of matrix $\mathbf{A}$.
$|\mathbf{A}|$ denotes the determinant of matrix $\mathbf{A}$. $\log (.)$ denotes the base-2 logarithm. 
 The notation $\mathbf{A}\preccurlyeq\mathbf{B}$ is
 equivalent to $\mathbf{B}-\mathbf{A}$ is a positive semi-definite
matrix. Motivated by the definition in \cite{zheng_tse}, we define the notation $f(P) \doteq g(P)$ as $\lim_{P \to \infty} \frac{f(P)}{\log (P)} = \lim_{P \to \infty} \frac{g(P)}{\log (P)}$. Similarly, $f(P) \dot \leq g(P)$ and $f(P) \dot \geq g(P)$ are equivalent to $\lim_{P \to \infty} \frac{f(P)}{\log (P)} \leq \lim_{P \to \infty} \frac{g(P)}{\log (P)}$ and $\lim_{P \to \infty} \frac{f(P)}{\log (P)} \geq \lim_{P \to \infty} \frac{g(P)}{\log (P)}$, respectively. Finally, we use $A \approx B$ to denote the approximate equality between $A$ and $B$, such that by substituting $A$ by $B$ the validity of the equations are not compromised.

\section{System Model}
Our setup consists of $K$ relays assisting the transmitter and the
receiver in the half-duplex mode, i.e. at a given time, the relays
can either transmit or receive. Each two nodes are assumed either i) to be
 connected by a quasi-static flat Rayleigh-fading channel,
i.e. the channel gains remain constant during a block of
transmission and change independently from block to block; or ii) to be
disconnected, i.e. there is no direct link between them. Hence, the
undirected graph $G=(V,E)$ is used to show the connected pairs in
the network\footnote{Note that however, in Remarks 2 and 6, the directed graph is considered.}. The node set is denoted by $V=\left\{0,1,\dots,{K+1}
\right\}$ where the $i$'th node is equipped with $N_i$ antennas.
Nodes $0$ and ${K+1}$ correspond to the transmitter and the receiver
nodes, respectively\footnote{Throughout the paper, it is assumed
that the network consists of one transmitter. However, in Theorems 5
and 6, we study the case of two-hop multiple transmitters single
receiver scenario.}. The received and the transmitted vectors at the
$k$'th node are shown by $\mathbf y_k$ and $\mathbf x_k$,
respectively. Hence, at the receiver side of the $a$'th node, we
have
\begin{equation}
\mathbf y_a = \sum_{\left\{a, b\right\} \in E} {\mathbf H_{a,b} \mathbf x_b} + \mathbf n_a,
\end{equation}
where $\mathbf H_{a,b}$ shows the $N_a \times N_b$
Rayleigh-distributed channel matrix between the $a$'th and the
$b$'th nodes and $\mathbf n_a \sim \mathcal N \left(\mathbf
0,\mathbf I_{N_a} \right)$ is the additive white Gaussian noise. We
assume reciprocal channels between each two nodes. Hence, $\mathbf
H_{a,b}=\mathbf H_{b,a}^T$. However, it can be easily verified that
all the statements of the paper are valid under the
non-reciprocity assumption. In the scenario of
single-antenna networks, the channel between nodes $a$ and $b$ is
denoted by $h_{\left\{a,b\right\}}$ to emphasize both the SISO and
the reciprocally assumptions. As in \cite{azarian} and
\cite{yang_belfiore2}, each relay is assumed to know the state of
its backward channel, and moreover, the receiver knows the
equivalent end-to-end channel. Hence, unlike the CF scheme in
\cite{yuksel}, no CSI feedback is needed. All nodes have the same
power constraint, $P$. Finally, we assume that the topology of the
network is known by the nodes such that they can perform a
distributed AF strategy throughout the network.

Throughout the section on diversity-multiplexing tradeoff, we make
some further assumptions in order to prove our statements. First, we
consider the scenario in which nodes with a single antenna are used.
Moreover, in Theorems 2, 3, 5, and 6, where we address DMT optimality
of the RS scheme,  we assume that there is no direct link between
the transmitter(s) and the receiver. This assumption is reasonable
when the transmitter and the receiver are far from each other and
the relay nodes establish the connection between the end nodes.
Moreover, we assume that all the relay nodes are connected to the
transmitter and to the receiver through quasi-static flat
Rayleigh-fading channels. Hence, the network graph is two-hop. In
specific, we denote the output vector at the transmitter as $\mathbf
x$, the input vector and the output vector at the $k$'th relay as
$\mathbf r_k$ and $\mathbf t_k$, respectively, and the input at the
receiver as $\mathbf y$.

\section{Proposed Random Sequential (RS) Amplify-and-Forwarding Scheme}

In the proposed RS scheme, a sequence $\mathrm P \equiv
\left(\mathrm p_1, \mathrm p_2, \dots, \mathrm p_L \right)$ of $L$
paths\footnote{Throughout the paper, a path  $\mathrm p$ is defined
as a sequence of the graph nodes $(v_0, v_1, v_2, \dots, v_l)$ such
that for any $i$, $\left\{v_i, v_{i+1} \right\} \in E$, and for all
$i \neq j$, we have $v_i \neq v_j$. The length of the path is
defined as the total number of edges on the path, $l$. Furthermore,
$\mathrm p(i)$ denotes the $i$'th node that $\mathrm p$ visits, i.e.
$\mathrm p(i) = v_i$.}  originating from the transmitter and
destinating to the receiver with the length $\left(l_1, l_2, \dots,
l_L \right)$ are involved in connecting the transmitter to the
receiver sequentially ($\mathrm p_i(0)=0, \mathrm p_i(l_i)=K+1$).
Note that any path $\mathrm p$ of $G$ can be selected multiple times
in the sequence.

Furthermore, the entire block of transmission is divided into $S$
slots, each consisting of $T'$ symbols. Hence, the entire block
consists of $T=ST'$ symbols. Let us assume the transmitter intends
to send information to the receiver at a rate of $r$ bits per
symbol. To transmit a message $w$, the transmitter selects the
corresponding codeword from a Gaussian random code-book consisting
of $2^{ST'r}$ elements each of with length $LT'$. Starting from the first
slot, the transmitter sequentially transmits the $i$'th portion ($1
\leq i \leq L$) of the codeword through the sequence of relay nodes
in $\mathrm p_i$. More precisely, a timing sequence $\left\{s_{i,j}
\right\}_{i=1,j=1}^{L,l_i}$ is associated with the path sequence.
The transmitter sends the $i$'th portion of the codeword in the
$s_{i,1}$'th slot. Following the transmission of the $i$'th portion of
the codeword by the transmitter, in the $s_{i,j}$'th slot, $1 \leq j
\leq l_i$, the node $\mathrm p_i(j)$ receives the transmitted signal
from the node $\mathrm p_i(j-1)$. Assuming $\mathrm p_i(j)$ is not
the receiver node, i.e. $j<l_i$,  it multiplies the received signal in the
$s_{i,j}$'th slot by a
$N_{\mathrm p_i(j)} \times N_{\mathrm p_i(j)}$ random, uniformly
distributed unitary matrix $\mathbf U_{i,j}$ which is known at the
receiver side, amplifies the signal by the maximum possible
coefficient $\alpha_{i,j}$ considering the output power constraint
$P$ and $\alpha_{i,j} \leq 1$, and transmits the amplified signal in
the $s_{i,j+1}$'th slot. Furthermore, the timing sequence $\left\{s_{i,j}
\right\}$ should have the following properties
\begin{enumerate}
 \item [(1)] for all $i,j$, we have $1 \leq s_{i,j} \leq S$.
 \item [(2)] for $i < i'$, we have $s_{i,1} < s_{i',1}$ (the ordering assumption on the paths)
 \item [(3)] for $j < j'$, we have   $s_{i,j} < s_{i,j'}$ (the causality assumption)
 \item [(4)] for all $i < i'$ and $s_{i,j} = s_{i',j'}$, we have
 $\left\{ \mathrm p_i(j), \mathrm p_{i'}(j'-1) \right\} \notin E$
 (no noncausal interference assumption). This assumption ensures that the signal of the
 future paths causes no interference on the output signal of the
current path. This assumption can be realized by designing the
timing of the paths such that in each time slot, the current running
paths are established through disjoint hops.
\end{enumerate}

At the receiver side, having received the signal of all paths, the
receiver decodes the transmitted message $w$ based on the signal
received in the time slots $\left\{s_{i,l_i}\right\}_{i=1}^L$. As we
observe in the sequel, the fourth assumption on
$\left\{s_{i,j}\right\}$ converts the equivalent end-to-end channel
matrix to lower-triangular in the case of single-antenna nodes, or
to block lower-triangular in the case of multiple-antenna nodes.

\begin{figure}[t]
  \centering
  \includegraphics[scale=0.8]{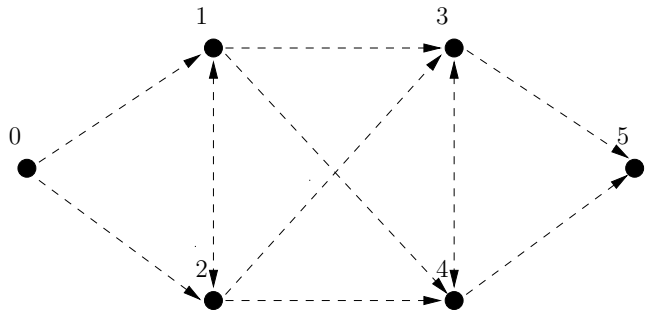}
\caption{An example of a 3 hops network where $N_0=N_5=2, N_1=N_2=N_3=N_4=1$.}
\label{fig:example}
\end{figure}

\begin{table}[b]
\begin{center}
\begin{tabular}{|c||c|c|c|c|c|c|c|}
\hline
time-slot & 1 & 2 & 3 & 4 & 5 & 6 & 7 \\
\hline \hline $\mathrm P_1(1)$ & $0 \to 1 $ &  $1 \to 3 $ & $3 \to 5 $ & --- & --- & --- & --- \\
\hline $\mathrm P_1(2)$ & --- & $0 \to 2$ & $2 \to 4$ & $4 \to 5$ & --- & --- & --- \\
\hline $\mathrm P_1(3)$ & --- & --- & --- & $0 \to 1$ & $1 \to 4$ & $4 \to 5$ & --- \\
\hline $\mathrm P_1(4)$ & --- & --- & --- &  --- & $0 \to 2$ & $2 \to 3$ & $3 \to 5$ \\
\hline
\end{tabular}
\caption{One possible valid timing for RS scheme with the path sequence
$\mathrm P_1= \left( \mathrm p_1, \mathrm p_2, \mathrm p_3, \mathrm p_4 \right)$.}
\label{tbl:1}
\end{center}
\end{table}

\begin{table}[t]
\begin{center}
\begin{tabular}{|c||c|c|c|c|c|c|}
\hline
time-slot & 1 & 2 & 3 & 4 & 5 & 6 \\
\hline \hline $\mathrm P_2(1)$ & $0 \to 1 $ &  $1 \to 3 $ & $3 \to 5 $ & --- & --- & --- \\
\hline $\mathrm P_2(2)$ & --- & $0 \to 2$ & $2 \to 4$ & $4 \to 5$ & --- & --- \\
\hline $\mathrm P_2(3)$ & --- & --- & $ 0 \to 1 $ &  $1 \to 3 $ & $3 \to 5 $ & ---\\
\hline $\mathrm P_2(4)$ & --- & --- & --- & $0 \to 2$ & $2 \to 4$ & $4 \to 5$ \\
\hline
\end{tabular}
\caption{One possible valid timing for RS scheme with the path sequence
$\mathrm P_2= \left( \mathrm p_1, \mathrm p_2, \mathrm p_1, \mathrm p_2 \right)$.}
\label{tbl:2}
\end{center}
\end{table}

An example of a three-hop network consisting of $K=4$ relays is
shown in figure (\ref{fig:example}). It can easily be verified that
there are exactly 12 paths in the graph connecting the transmitter
to the receiver. Now, consider the four paths $\mathrm
p_1=(0,1,3,5)$, $\mathrm p_2=(0,2,4,5)$, $\mathrm p_3=(0,1,4,5)$ and
$\mathrm p_4=(0,2,3,5)$ connecting the transmitter to the receiver.
Assume the RS scheme is performed with the path sequence $\mathrm
P_1 \equiv (\mathrm p_1, \mathrm p_2, \mathrm p_3, \mathrm p_4)$.
Table \ref{tbl:1} shows one possible valid timing sequence
associated with RS scheme with the path sequence $\mathrm P_1$. As
seen, the first portion of the transmitter's codeword is sent in the
$1$st time slot and is received by the receiver through the nodes of
the path $\mathrm P_1(1)\equiv(0, 1, 3, 5)$ as follows: In the $1$st
slot, the transmitter's signal is received by node $1$. Following
that, in the $2$nd slot, node $1$ sends the amplified signal to node
$3$, and finally, in the $3$rd slot, the receiver receives the
signal from node $3$. As observed, for every $1 \leq i \leq 3$, signal
of the $i$'th path interferes on the output signal of the $i+1$'th
path. However, no interference is caused by the signal of future
paths on the outputs of the current path. The timing sequence
corresponding to Table \ref{tbl:1} can be expressed as
$s_{i,j}=i+\lfloor \frac{i}{3}\rfloor+j-1$ and it results in the
total number of transmission slots to be equal to $7$, i.e. $S=7$.

As an another example, consider RS scheme with the path sequence
$\mathrm P_2 \equiv (\mathrm p_1, \mathrm p_2, \mathrm p_1, \mathrm
p_2)$. Table \ref{tbl:2} shows one possible valid timing-sequence
for the RS scheme with the path sequence $\mathrm P_2$. Here, we
observe that the signal on every path interferes on the output of
the next two consecutive paths. However, like the scenario with
$\mathrm P_1$, no interference is caused by the signal of future
paths on the output signal of the current path. The timing sequence
corresponding to Table \ref{tbl:2} can be expressed as
$s_{i,j}=i+j-1$ and it results in the total number of transmission
slots equal to $6$, i.e. $S=6$.

It is worth noting that to achieve higher spectral efficiencies
(corresponding to larger  multiplexing gains), it is desirable to
have larger values for $\frac{L}{S}$. Indeed, $\frac{L}{S} \to 1$ is
the highest possible value. However, this can not be achieved in
some graphs (an example is the case of two-hop single relay scenario
studied in the next section where $\frac{L}{S}=0.5$). On the other
hand, to achieve higher reliability (corresponding to larger
diversity gains between the end nodes), it is desirable to utilize
more paths of the graph in the path sequence. It is not always
possible to satisfy both of these objectives simultaneously. As an
example, consider the single-antenna two-hop relay network where
there is a direct link between the end nodes, i.e. $G$ is the
complete graph. Here, all the nodes of the graph interfere on each
other,  and consequently, in each time slot only one path can transmit signal. Hence,
in order to achieve $\frac{L}{S} \to 1$, only the direct path $(0,
K+1)$ should be utilized for almost all the time.

As an another example, consider the 3-hop network in figure
(\ref{fig:example}). As  we will see in the following sections, the
RS scheme corresponding to the path sequence $\mathrm P_1$ achieves
the maximum diversity gain of the network, $d=4$. However, it can
easily be verified that no valid timing-sequence can achieve fewer
number of transmission slots than the one shown in Table
\ref{tbl:1}. Hence, $\frac{L}{S}=\frac{4}{7}$ is the best RS scheme
can achieve with $\mathrm P_1$. On the other hand, consider the RS
scheme with the path sequence $\mathrm P_2$. Although, as seen in
the sequel, the scheme achieves the diversity gain $d=2$ which is
below the maximum diversity gain of the network, it utilizes fewer
number of slots compared to the case using the path sequence
$\mathrm P_1$. Indeed, it achieves $\frac{L}{S}=\frac{4}{6}$.

In the two-hop scenario investigated in the next section, we will
see that for asymptotically large values of $L$, it is possible to
utilize all the paths needed to achieve the maximum diversity gain
and, at the same time, devise the timing sequence such that
$\frac{L}{S} \to 1$. Consequently, it will be shown that in this
setup, the proposed RS scheme achieves the optimum DMT.

\section{Diversity-Multiplexing Tradeoff}
In this section, we analyze the performance of the RS scheme in
terms of the DMT for the single-antenna multiple-relay networks.
First, in subsection \textit{A}, we study the performance of the RS
scheme for the case of non-interfering relays where there exists
neither causal nor noncausal interference between the signals sent
through different paths. In this case, as there exists no
interference between different paths, we can assume that the
amplification coefficients take values greater than one, i.e. the
constraint $\alpha_{i,j} \leq 1$ can be omitted. Under the condition
of non-interfering relays, we derive the exact DMT of the RS scheme.
As a result, we show that the RS scheme achieves the optimum DMT for
the setup of non-interfering two-hop multiple-relay ($K>1$)
single-transmitter single-receiver, where there exists no direct
link between the relay nodes and between the transmitter and the
receiver (more precisely, $E=\left\{\left\{ 0, k \right\},\left\{ k,
K+1 \right\} \right\}_{k=1}^K$). To prove this, we assume that the
RS scheme relies on $L=BK$ paths, $S=BK+1$ slots, where $B$ is an integer number, and the path
sequence is $\mathrm Q \equiv \left(\mathrm q_1, \dots, \mathrm q_K,
\mathrm q_1, \dots, \mathrm q_K, \dots, \mathrm q_1, \dots,
\mathrm q_K\right)$ where $\mathrm q_k\equiv(0, k, K+1)$. In other
words, every path $\mathrm q_k$ is used $B$ times in the sequence.
Here, each $K$ consecutive slots are called a sub-block. Hence, the
entire block of transmission consists of $B+1$ sub-blocks. The
timing sequence is defined as $s_{i,j}=i+j-1$. It is easy to verify
that the timing sequence satisfies the requirements. Here, we
observe that the spectral efficiency is $\frac{L}{S}=1-\frac{1}{S}$
which converges to 1 for asymptotically large values of $S$. By
deriving the exact DMT of the RS scheme, we prove that the RS scheme
achieves the optimum DMT for asymptotically large values of $S$.

In subsection \textit{B}, we study the performance of the RS scheme
for general single-antenna multiple-relay networks. First, we study
the performance of RS scheme for the setup of two-hop
single-transmitter single-receiver multiple-relay ($K > 1$) networks
where there exists no direct link between the transmitter and the
receiver; However, no additional restriction is imposed on the graph
of the interfering relay pairs. We apply the RS scheme with the same
parameters used in the case of two-hop non-interfering networks. We
derive a lower-bound for DMT of the RS scheme. Interestingly, it
turns out that the derived lower-bound merges to the upper-bound on
the DMT for asymptotic values of $B$. Next, we generalize our result
and derive a lower-bound on DMT of the RS scheme for general
single-antenna multiple-relay networks.

Finally, in subsection \textit{C}, we generalize our results for the
scenario of single-antenna two-hop multiple-access multiple-relay
($K>1$) networks where there exists no direct link between the
transmitters and the receiver. Here, we apply the RS scheme with the
same parameters as used in the case of single-transmitter
single-receiver two-hop relay networks. However, it should be noted
that here, instead of sending data from the single transmitter, all
the transmitters send data coherently. By deriving a lower-bound on
the DMT of the RS scheme, we show that in this network the RS scheme
achieves the optimum DMT. However, as studied in subsection
\textit{D}, for the setup of single-antenna two-hop multiple-access
single-relay networks where there exists no direct link between the
transmitters and the receiver, the proposed RS scheme reduces to
naive amplify-and-forward relaying and is not optimum in terms of
the DMT. In this setup, we show that the DDF scheme achieves
the optimum DMT.

\subsection{Non-Interfering Relays}

In this subsection, we study the DMT behavior of the RS scheme in
general single-antenna multi-hop relay networks under the condition
that there exists neither causal nor noncausal interference between
the signals transmitted over different paths. More precisely, we
assume the timing sequence is designed such that if
$s_{i,j}=s_{i',j'}$, then we have $\left\{ \mathrm p_i(j), \mathrm
p_{i'}(j'-1) \right\} \notin E$. This assumption is stronger than
the fourth assumption on the timing sequence (here the condition $i
< i'$ is omitted). We call this the ``non-interfering relaying''
condition. Under this condition, as there exists no interference
between signals over different paths, we can assume that the
amplification coefficients take values greater than one, i.e. the
constraint $\alpha_{i,j} \leq 1$ can be omitted.

First, we need the following definition.
\begin{definition}
For a network with the connectivity graph $G=(V,E)$, a cut-set on
$G$ is defined as a subset $\mathcal S \subseteq V$ such that $0 \in
\mathcal S, K+1 \in \mathcal S^c$. The weight of the cut-set
corresponding to $\mathcal S$, denoted by $w(\mathcal S)$, is
defined as
\begin{eqnarray}
w_G(\mathcal S) & = & \sum_{a \in \mathcal S, b \in \mathcal S^c, \left\{a, b \right\} \in E}{N_a \times N_b}.
\end{eqnarray}
\end{definition}

\begin{thm}
Consider a half-duplex single-antenna multiple-relay network with
the connectivity graph $G=(V,E)$. Assuming ``non-interfering
relaying'', the RS scheme with the path sequence $\left(\mathrm p_1, \mathrm p_2, \dots ,\mathrm p_L \right)$ achieves the diversity
gain corresponding to the following linear programming optimization problem
\begin{equation}
d_{RS, NI}(r) = \displaystyle \min_{\boldsymbol \mu \in \mathcal {\hat R}} \sum_{e \in E} \mu_e, \label{eq:t1_exact}
\end{equation}
where $\boldsymbol \mu$ is a vector defined on edges of $G$ and
$\mathcal {\hat R}$ is a region of $\boldsymbol \mu$ defined as
$$\mathcal {\hat R} \equiv \left\{ \boldsymbol \mu \left| \, \,
\mathbf 0 \leq \boldsymbol \mu \leq \mathbf 1, \displaystyle
\sum_{i=1}^{L} \max_{1 \leq j \leq l_i} \mu_{\left\{ \mathrm p_i(j),
\mathrm p_i(j-1) \right\}} \geq L - Sr \right\} \right..$$ Furthermore, the DMT
of the RS scheme can be upper-bounded as
\begin{equation}
d_{RS, NI}(r) \leq (1-r)^+\min_{\mathcal S} w_G( \mathcal S), \label{eq:t1_ub}
\end{equation}
where $\mathcal S$ is a cut-set on $G$. Finally, by properly
selecting the path sequence, one can always achieve
\begin{equation}
d_{RS, NI}(r) \geq \left(1-l_G r\right)^+ \min_{\mathcal S} w_G(\mathcal S), \label{eq:t1_lb}
\end{equation}
where $\mathcal S$ is a cut-set on $G$ and $l_G$ is the maximum path length between  the transmitter and the receiver.
\end{thm}
\begin{proof}
Since the relay nodes are non-interfering, the achievable rate of
the RS scheme for a realization of the channels is equal to
\begin{IEEEeqnarray}{l}
R_{RS,NI}\left(\left\{ h_e\right\}_{e \in E}\right) = \nonumber \\
\frac{1}{S} \sum_{i=1}^{L} \log \left( 1 + P \prod_{j=1}^{l_i}\left|
\alpha_{i,j} \right|^2 \left| h_{\left\{\mathrm p_i(j), \mathrm p_i(j-1) \right\}}
\right|^2 \left(1 + \sum_{j=1}^{l_i-1} \prod_{k=j}^{l_i-1}{\left| \alpha_{i,k} \right|^2 \left|
h_{\left\{\mathrm p_i(k), \mathrm p_i(k+1) \right\}} \right|^2} \right)^{-1} \right),
\end{IEEEeqnarray}
where $\forall j < l_i: \alpha_{i,j} =  \sqrt{\frac {P}{1+\left|
h_{\left\{\mathrm p_i(j-1), \mathrm p_i(j)\right\}} \right|^2P} }$
and $\alpha_{i, l_i}=1$ (since $\mathrm p_i(l_i)=K+1$). In deriving the above equation, we have used the fact that as the paths are non-interfering, the achievable rate can be written as the sum of the rates over the paths, noting that the terms  $P \prod_{j=1}^{l_i}\left|
\alpha_{i,j} \right|^2 \left| h_{\left\{\mathrm p_i(j), \mathrm p_i(j-1) \right\}}
\right|^2$ and $1 + \sum_{j=1}^{l_i-1} \prod_{k=j}^{l_i-1}{\left| \alpha_{i,k} \right|^2 \left|
h_{\left\{\mathrm p_i(k), \mathrm p_i(k+1) \right\}} \right|^2}$ represent the effective signal power and the noise power over the $i$th path, respectively.
Hence, the probability of outage equals
\begin{eqnarray} \label{pout}
\mathbb P\left\{ \mathcal E \right\} & = & \mathbb P \left\{ R_{RS,NI}\left(\left\{ h_e\right\}_{e \in E}\right) \leq r \log \left( P \right) \right\} \nonumber \\
& \stackrel{(a)}{\doteq} & \mathbb P \left\{ \prod_{i=1}^{L} \max \left\{ P^{-1} ,  \min \left\{ \left| h_{\left\{0, \mathrm p_i(1) \right\}} \right|^2 \prod_{k=1}^{j}{\left| \alpha_{i,k} \right|^2 \left| h_{\left\{\mathrm p_i(k), \mathrm p_i(k+1) \right\}} \right|^2} \right\}_{j=0}^{l_i-1} \right\} \leq P^{Sr - L} \right\} \nonumber \\
& \stackrel{(b)}{\doteq} & \displaystyle \max_{\begin{subarray}{c} t_1, t_2, \dots, t_L \\ 1 \leq t_i \leq l_i \end{subarray}
} \mathbb P \left\{ \prod_{i=1}^{L} \max \left\{P^{-1},  \left| h_{\left\{0, \mathrm p_i(1) \right\}} \right|^2 \prod_{k=1}^{t_i-1}{\left| \alpha_{i,k} \right|^2 \left| h_{\left\{\mathrm p_i(k), \mathrm p_i(k+1) \right\}} \right|^2} \right\} \leq P^{Sr - L} \right\} \nonumber \\
&  \stackrel{(c)}{\doteq} & \displaystyle \max_{\begin{subarray}{c} \mathcal S_1, \mathcal S_2, \dots, \mathcal S_L \\ \mathcal S_i \subseteq \left\{1,2,\dots, l_i-1 \right\} \end{subarray}
} \displaystyle \max_{\begin{subarray}{c} t_1, t_2, \dots, t_L \\ \max \left\{x \in \mathcal S_i\right\}  < t_i \leq l_i \end{subarray}
} \nonumber \\
&&\mathbb P \left\{ \prod_{i=1}^{L} \max \left\{P^{-1},  P^{\left| \mathcal S_i \right|}\left| h_{\left\{\mathrm p_i(t_i), \mathrm p_i(t_i-1) \right\}} \right|^2 \prod_{k \in \mathcal S_i}{ \left| h_{\left\{\mathrm p_i(k), \mathrm p_i(k-1) \right\}} \right|^2} \right\} \leq P^{Sr - L} \right\}. \label{eq:t1_ni_1}
\end{eqnarray}
Here, $(a)$ follows from the facts that i)~$\forall x\geq 0: \max
\left\{ 1, x \right\} \leq 1+x \leq 2 \max \left\{ 1, x \right\}$, which implies that $1+P \Theta \approx \max (1, P \Theta)$, where $$  \Theta \triangleq  \prod_{j=1}^{l_i}\left|
\alpha_{i,j} \right|^2 \left| h_{\left\{\mathrm p_i(j), \mathrm p_i(j-1) \right\}}
\right|^2 \left(1 + \sum_{j=1}^{l_i-1} \prod_{k=j}^{l_i-1}{\left| \alpha_{i,k} \right|^2 \left|
h_{\left\{\mathrm p_i(k), \mathrm p_i(k+1) \right\}} \right|^2} \right)^{-1},$$
and ii)~for all $x_i \geq 0$,  $\frac{1}{M} \min \left\{ \frac{1}{x_i}
\right\}_{i=1}^{M} \leq \left( \sum_{i=1}^{M} x_i\right)^{-1} \leq
\min \left\{ \frac{1}{x_i} \right\}_{i=1}^{M}$, which implies that $$  \left(1 + \sum_{j=1}^{l_i-1} \prod_{k=j}^{l_i-1}{\left| \alpha_{i,k} \right|^2 \left|
h_{\left\{\mathrm p_i(k), \mathrm p_i(k+1) \right\}} \right|^2} \right)^{-1} \approx \min \left(1, \left\{ \left(\prod_{k=j}^{l_i-1}{\left| \alpha_{i,k} \right|^2 \left|
h_{\left\{\mathrm p_i(k), \mathrm p_i(k+1) \right\}} \right|^2} \right)^{-1}\right\}_{j=1}^{l_i-1}\right).$$ $(b)$ follows from
the fact that for any increasing function $f(.)$ , we have
$$ \displaystyle \max_{1 \leq i \leq M} \mathbb P \left\{ f\left( x_i
\right) \leq y \right\} \leq \mathbb P \left\{ f\left( \displaystyle
\min_{1 \leq i \leq M} x_i \right) \leq y \right\} \leq
\displaystyle M \max_{1 \leq i \leq M} \mathbb P \left\{ f\left( x_i
\right) \leq y \right\} .$$ $(c)$ follows from the fact that $$0.5
\min \left\{1, P \left| h_{\left\{\mathrm p_i(k), \mathrm p_i(k-1)
\right\}} \right|^2 \right\} \leq \left| \alpha_{i,k}
h_{\left\{\mathrm p_i(k), \mathrm p_i(k-1) \right\}} \right|^2
\leq \min \left\{1, P \left| h_{\left\{\mathrm p_i(k), \mathrm p_i(k-1) \right\}} \right|^2 \right\} ,$$ which implies that $\left| \alpha_{i,k}
h_{\left\{\mathrm p_i(k), \mathrm p_i(k-1) \right\}} \right|^2
\leq \min \left\{1, P \left| h_{\left\{\mathrm p_i(k), \mathrm p_i(k-1) \right\}} \right|^2 \right\}$. In the last line of (\ref{pout}), $\mathcal{S}_i$ denotes the subset of $\{1,2, \cdots, t_i-1\}$ for which $ P \left| h_{\left\{\mathrm p_i(k), \mathrm p_i(k-1) \right\}} \right|^2 \leq 1$.

 Assuming $\left| h_e
\right|^2=P^{-\mu_e}$, we define the region $\mathcal R \subseteq
\mathbb R^{\left| E \right|}$ as the set of points $\boldsymbol \mu
= [\mu_e]_{e \in E}$ that the outage event occurs. Let us define
$\mathcal R_+ = \mathcal R \cap \left( \mathbb R_+ \cup \left\{ 0
\right\}\right)^{\left| E \right|}$. As the probability density
function diminishes exponentially as $e^{-P^{\mu_e}}$ for positive
values of $\mu_e$, we have $\mathbb P \left\{ \mathcal R_+ \right\}
\doteq \mathbb P \left\{ \mathcal R \right\}$. Hence, we have
\begin{eqnarray}
\mathbb P \left\{ \mathcal E \right\} & \doteq & \mathbb P \left\{ \mathcal R_+ \right\} \nonumber \\
& \stackrel{(a)}{\doteq} & \displaystyle \max_{\begin{subarray}{c} \mathcal S_1,
\mathcal S_2, \dots, \mathcal S_L \\ \mathcal S_i \subseteq \left\{1,2,\dots, l_i-1 \right\} \end{subarray}
} \displaystyle \max_{\begin{subarray}{c} t_1, t_2, \dots, t_L \\ \max \left\{x \in \mathcal S_i\right\}  < t_i \leq l_i \end{subarray}
}
\mathbb P \left\{ \mathcal R\left( \boldsymbol{\mathcal S} , \mathbf t \right)  \right\} \nonumber \\
& \stackrel{(b)}{\doteq} & \displaystyle \displaystyle \max_{\begin{subarray}{c} \mathbf t \\ 1 \leq t_i \leq l_i \end{subarray}
}
\mathbb P \left\{ \mathcal R_0\left( \mathbf t \right)  \right\}, \label{eq:hd_show}
\end{eqnarray}
where
$$\mathcal R\left( \boldsymbol{\mathcal S}, \mathbf t \right) \equiv \left\{ \boldsymbol \mu \in \left( \mathbb R_+ \cup
\left\{ 0 \right\}\right)^{\left| E \right|} \left| \sum_{i=1}^{L}
\min \left\{1, \mu_{\left\{ \mathrm p_i(t_i), \mathrm p_i(t_i-1)
\right\}} + \displaystyle \sum_{k \in \mathcal S_i} \mu_{\left\{
\mathrm p_i(k), \mathrm p_i(k-1) \right\}} - \left| \mathcal S_i
\right|  \right\} \geq L - Sr  \right\} \right.,$$ $\mathbf t = \left[ t_1,
t_2, \dots, t_L\right]$, $\boldsymbol{\mathcal{S}}= \left[ \mathcal{S}_1, \cdots, \mathcal{S}_L\right]$, and $\mathcal R_0\left( \mathbf t \right)
\equiv \mathcal R\left( \oslash , \oslash, \dots, \oslash, t_1, t_2,
\dots, t_L \right)$, in which $\oslash$ denotes the null set. Here, $(a)$ follows from \eqref{eq:t1_ni_1}. In order to prove $(b)$, we first show that
\begin{equation}
\min \left\{1, \mu_{\left\{ \mathrm p_i(t_i), \mathrm p_i(t_i-1) \right\}} + \displaystyle \sum_{k \in
\mathcal S_i} \mu_{\left\{ \mathrm p_i(k), \mathrm p_i(k-1)
\right\}}  - \left| \mathcal S_i \right|  \right\} \leq
\displaystyle \max_{t'_i \in \mathcal S_i \cup \left\{ t_i \right\}}
\min \left\{1,  \mu_{\left\{ \mathrm p_i(t'_i), \mathrm p_i(t'_i-1) \right\}} \right\}. \label{eq:easy_show}
\end{equation}

In order to verify \eqref{eq:easy_show}, consider two possible
scenarios: i) for all $t'_i \in \mathcal S_i \cup \left\{ t_i
\right\}$, we have $\mu_{\left\{ \mathrm p_i(t'_i), \mathrm
p_i(t'_i-1) \right\}} \leq 1$. In this scenario, as in the left hand
side of the inequality, we have the summation of $| \mathcal S_i | +
1$ positive parameters with value less than or equal to $1$
subtracted by $|\mathcal S_i|$, we conclude that the left hand side of
the inequality is less than or equal to $\mu_{\left\{ \mathrm
p_i(t'_i), \mathrm p_i(t'_i-1) \right\}}$ for any $t' \in \mathcal
S_i \cup \left\{ t_i \right\}$. Hence, \eqref{eq:easy_show} is
valid; ii) At least for one $t' \in \mathcal S_i \cup \left\{ t_i
\right\}$, we have $\mu_{\left\{ \mathrm p_i(t'_i), \mathrm
p_i(t'_i-1) \right\}} > 1$. In this scenario, the right hand side of
the inequality is equal to $1$ and accordingly,
\eqref{eq:easy_show} is valid. According to \eqref{eq:easy_show}, we
have $ \mathcal R \left(\boldsymbol{\mathcal S}_1, \mathbf{t} \right) \subseteq \displaystyle
\bigcup_{\begin{subarray}{c} \mathbf t' \\ t'_i \in \mathcal S_i
\cup \left\{ t_i \right\} \end{subarray}} \mathcal R_0\left( \mathbf
t' \right)$, which results in $(b)$ of \eqref{eq:hd_show}.

On the other hand, we know that for $\boldsymbol \mu^0 \geq \mathbf
0$, we have $\mathbb{P} \left\{\boldsymbol \mu \geq \boldsymbol
\mu^0 \right\} \doteq P^{-\mathbf{1} \cdot  \boldsymbol \mu^0}$. By
taking derivative with respect to $\boldsymbol \mu$, we have
$f_{\boldsymbol \mu}(\boldsymbol \mu) \doteq P^{- \mathbf 1 \cdot
\boldsymbol \mu}$. Let us define  $l_0 \triangleq
\displaystyle \min_{\boldsymbol \mu \in \mathcal R_0(\mathbf t)}
\mathbf 1 \cdot \boldsymbol \mu$ and $\boldsymbol \mu_0 \triangleq
\displaystyle \arg \, \min_{\boldsymbol \mu \in \mathcal
R_0(\mathbf t)} \mathbf 1 \cdot \boldsymbol \mu$,  $\mathcal I \triangleq \left[0, l_0 \right]^{2K}$, $\mathcal
I_0^c \triangleq [\mu_0(1), \infty ) \times [\mu_0(2), \infty ) \times \dots
\times [\mu_0(L), \infty )$ and for $1 \leq i \leq L$,
$\mathcal{I}_i^c \triangleq [0, \infty )^{i-1} \times [l_0, \infty ) \times [0,
\infty )^{L-i}$. It is easy to verify that $\mathcal I_0^c \subseteq
\mathcal R_0(\mathbf t)$. Hence, we have
\begin{eqnarray}
\mathbb{P} \left\{ \mathcal R_0(\mathbf t) \right\} & \stackrel{(a)}{\doteq}
& \mathbb{P} \left\{  \mathcal I_0^c \right\} +  \int_{R_0(\mathbf t) \bigcap \mathcal{I}}{f_{\boldsymbol
\mu}\left(\boldsymbol \mu \right) d\boldsymbol \mu
} + \sum_{i=1}^{L}{\mathbb{P} \left\{ \mathcal R_0( \mathbf t ) \cap \mathcal{I}_i^c \right\}}
\nonumber
\\
&\stackrel{(b)}{\doteq} & P^{-l_0}.
\end{eqnarray}
Here, $(a)$ follows from the facts that i)~$\mathbb P \left\{
\bigcup_{i=1}^M \mathcal A_i \right\} \doteq \sum_{i=1}^{M} \mathbb
P \left\{ \mathcal A_i \right\}$, and ii)~$\mathcal I_0^c \subseteq \mathcal
R_0( \mathbf t)$ and $\mathbb R_+^L = \mathcal I \bigcup \left(
\bigcup_{i=1}^{L} \mathcal I_i^c \right)$ which imply that $\mathcal R_0(\mathbf t) $ can be written as $\mathcal I_0^c \bigcup \left( \mathcal R_0(\mathbf t) \bigcap \mathcal I\right) \bigcup \left[ \bigcup_{i=1}^M \left( \mathcal R_0(\mathbf t) \bigcap   \mathcal I_i^c\right)\right]$. $(b)$ follows from the
facts that $\mathbb{P} \left\{  \mathcal I_0^c \right\} = \mathbb P
\left\{ \boldsymbol \mu \geq \boldsymbol \mu_0 \right\} \doteq
P^{-l_0}$, $\int_{R_0( \mathbf t) \bigcap
\mathcal{I}}{f_{\boldsymbol \mu}\left(\boldsymbol \mu \right)
d\boldsymbol \mu } \dot  \leq  \mbox{vol} \left(R_0( \mathbf t)
\bigcap \mathcal{I}\right) P^{-l_0}$, noting that $\mbox{vol} \left(R_0( \mathbf t)
\bigcap \mathcal{I}\right)$ is a constant number independent of $P$, and $\mathbb{P} \left\{
\mathcal R_0( \mathbf t ) \cap \mathcal{I}_i^c \right\} \leq
\mathbb{P} \left\{ \mathcal{I}_i^c \right\} = P^{-l_0}$. Now,
defining $g_{\mathbf t} (\boldsymbol \mu)=\sum_{i=1}^{L} \min
\left\{1, \mu_{\left\{ \mathrm p_i(t_i), \mathrm p_i(t_i-1)
\right\}} \right\}$ and $ \boldsymbol {\hat \mu} = [\min \left\{
\mu_e, 1\right\}]_{e \in E}$, it is easy to verify that $g_{\mathbf
t}(\boldsymbol {\hat \mu} ) = g_{\mathbf t}(\boldsymbol  \mu )$ and
at the same time $\mathbf 1 \cdot \boldsymbol {\hat \mu} < \mathbf 1
\cdot \boldsymbol \mu$ unless $\boldsymbol {\hat \mu} = \boldsymbol
\mu$. Hence, defining $\hat g_{\mathbf t}(\boldsymbol \mu) =
\sum_{i=1}^{L} \mu_{\left\{ \mathrm p_i(t_i), \mathrm p_i(t_i-1)
\right\}} $, we have
\begin{equation}
d_{RS, NI}(r) = \min_{\begin{subarray}{c}
                       \mathbf t \\
            1 \leq t_i \leq l_i
                      \end{subarray}
} \min_{\begin{subarray}{c}
         \boldsymbol \mu \geq \mathbf 0 \\
    g_{\mathbf t} (\boldsymbol \mu) \geq L - Sr
        \end{subarray}
} \mathbf 1 \cdot \boldsymbol \mu = \min_{\begin{subarray}{c}
                       \mathbf t \\
            1 \leq t_i \leq l_i
                      \end{subarray}
} \min_{\begin{subarray}{c}
         \mathbf 0 \leq \boldsymbol \mu \leq \mathbf 1 \\
     \hat g_{\mathbf t}(\boldsymbol \mu) \geq L - Sr
        \end{subarray}
} \mathbf 1 \cdot \boldsymbol \mu = \min_{\boldsymbol \mu \in \mathcal {\hat R}} \, \mathbf 1 \cdot \boldsymbol \mu,
\end{equation}
where $\mathcal {\hat R} = \left\{ \boldsymbol \mu \left| \, 
\mathbf 0 \leq \boldsymbol \mu \leq \mathbf 1, \displaystyle
\sum_{i=1}^{L} \max_{1 \leq j \leq l_i} \mu_{\left\{ \mathrm p_i(j),
\mathrm p_i(j-1) \right\}} \geq L - Sr \right\} \right.$. This proves the first part of the theorem.

Now, let us define $G_{\mathrm P}=(V, E_{\mathrm P})$ as the
subgraph of $G$ consisting of the edges in the path sequence, i.e.
$E_{\mathrm P}=\left\{ \left\{ \mathrm p_i(j), \mathrm p_i(j-1)
\right\}, \forall i,j: 1 \leq i \leq L, 1 \leq j \leq l_i \right\}$.
Assume $\mathcal {\hat S} = \displaystyle \underset{\mathcal
S}{\mbox{argmin}} \, w_{G_{\mathrm P}}(\mathcal S), $ where $\mathcal
S$ is a cut-set on $G_{\mathrm P}$. We define $\boldsymbol { \hat
\mu}$ as $ \hat \mu_e = \frac{(L-Sr)^+}{L}$ for all $e
\in E_{\mathrm P}$ such that $|e \cap \mathcal{\hat S}|=|e \cap
\mathcal{\hat S}^c| = 1$ and $ \hat \mu_e = 0$ for the
other edges $e \in E$. As all the paths cross the cutset $\mathcal{\hat S}$ at least once, it follows that $\max_{1 \leq j \leq l_i} \mu_{\left\{ \mathrm p_i(j),
\mathrm p_i(j-1) \right\}} = \frac{(L-Sr)^+}{L}$, which implies that $\boldsymbol {\hat
\mu} \in \mathcal {\hat R}$. Hence, we have
\begin{equation}
d_{RS,NI}(r) \leq \mathbf 1 \cdot \boldsymbol {\hat \mu} = \frac{(L-Sr)^+}{L}
\min_{\mathcal S} w_{G_{\mathrm P}}(\mathcal S) \stackrel{(a)}{\leq} \frac{(L-Sr)^+}{L}
\min_{\mathcal S} w_{G}(\mathcal S) \stackrel{(b)}{\leq} (1-r)^+ \min_{\mathcal S} w_{G}(\mathcal S),
\end{equation}
where $(a)$ follows from the fact that as $G_{\mathrm P}$ is a sub-graph of $G$, we have $\min_{\mathcal S} \, w_{G_{\mathrm P}}(\mathcal S)  \leq \min_{\mathcal S} \, w_{G}(\mathcal S) $ and $(b)$ results from $S \geq L$. This proves the second part of the Theorem.

Finally, we prove the lower-bound on the DMT of the RS scheme. Let
us define $d_G=\min_{\mathcal S} w_G (\mathcal S)$. Consider the
maximum flow algorithm \cite{graph_book} on $G$  from the source
node $0$ to the sink node $K+1$. According to the Ford-Fulkerson
Theorem \cite{graph_book}, one can achieve the maximum flow which is
equal to the minimum cut of $G$ by the union of elements of a
sequence $\left(\mathrm  {\hat p}_1, \mathrm {\hat p}_2, \dots,
\mathrm {\hat p}_{d_G} \right)$ of paths with the lengths
$\left(\hat l_1,\hat l_2, \dots, \hat l_{d_G}\right)$. Now, consider
the RS scheme with $L=L_0 d_G$ paths and the path sequence
$\left(\mathrm p_1, \mathrm p_2, \dots, \mathrm p_L\right)$
consisting of the paths that achieve the maximum flow of $G$ such
that any path $\mathrm {\hat p}_i$ occurs exactly $L_0$ times in the
sequence. Considering $\left(l_1, l_2, \dots, l_L\right)$ as the
length sequence, we select the timing sequence as
$s_{i,j}=\sum_{k=1}^{i-1}l_k+j$. It is easy to verify that, not
only the timing sequence satisfies the 4 requirements needed for the
RS scheme, but also the active relays with the timing sequence are
non-interfering.  Hence, the assumptions of the first part of the
theorem are valid. Moreover, we have $S \leq l_{G} L$. According to
\eqref{eq:t1_exact}, the diversity gain of the RS scheme equals
\begin{equation}
d_{RS, NI}(r)=\min_{\boldsymbol \mu \in \mathcal {\hat R}} \sum_{e \in E} \mu_e. \label{eq:t1_ni_2}
\end{equation}
As $\boldsymbol \mu \in \mathcal {\hat R}$, we have
\begin{equation}
(L-Sr)^+ \leq \sum_{i=1}^{L} \max_{1 \leq j \leq l_i} \mu_{\left\{ \mathrm p_i(j),
\mathrm p_i(j-1) \right\}} \stackrel{(a)}{\leq} L_0 \sum_{e \in E} \mu_e,\label{eq:t1_ni_3}
\end{equation}
where $(a)$ results from the fact that as $\left(\mathrm  {\hat
p}_1, \mathrm {\hat p}_2, \dots, \mathrm {\hat p}_{d_G} \right)$
form a valid flow on $G$ (they are non-intersecting over $E$), every $e \in E$ occurs in at most one
$\mathrm {\hat p}_i$, or equivalently, in at most $L_0$ number of
$\mathrm p_i$'s. Combining \eqref{eq:t1_ni_2} and
\eqref{eq:t1_ni_3}, we have
\begin{equation}
d_{RS, NI}(r) \geq \frac{(L-Sr)^+}{L_0} \geq \left(1-l_G r\right)^+d_G = \left(1-l_G r\right)^+ \min_{\mathcal S} w_G(\mathcal S). \label{eq:t1_lb_proof}
\end{equation}
This proves the third part of the Theorem.
\end{proof}

\textit{Remark 1-} In scenarios where the minimum-cut on $G$ is
achieved by a cut of the MISO or SIMO form, i.e., the edges that
cross the cut are either originated from or destined to the same vertex,
the upper-bound on the diversity gain of the RS scheme derived in
\eqref{eq:t1_ub} meets the information-theoretic upper-bound on the
diversity gain of the network. Hence, in this scenario, any RS
scheme that achieves \eqref{eq:t1_ub} indeed achieves the optimum
DMT.

\textit{Remark 2-} In general, the upper-bound \eqref{eq:t1_ub} can be achieved for various certain graph topologies by wisely designing the path sequence and the timing sequence. One example is the case of the layered network\cite{avesti_wireless_deterministic} in which all the paths from the source to the destination have the same length  $l_{G}$. Let us assume that the relays are allowed to operate in the full-duplex manner. In this case, it easily can be observed that the timing sequence corresponding to the path sequence $\left(\mathrm p_1, \mathrm p_2, \dots, \mathrm p_L\right)$ used in the proof of \eqref{eq:t1_lb} can be modified to $s_{i,j}=i+j-1$. Accordingly, the number of slots is decreased to $S=L+l_G-1$. Rewriting \eqref{eq:t1_lb_proof}, we have $d_{RS, NI}(r)=\left(1-r-\frac{l_G-1}{L}r\right)^+  \min_{\mathcal S} w_G(\mathcal S)$ which achieves $\left(1-r\right)^+  \min_{\mathcal S} w_G(\mathcal S)$ for large values of $L$.

Next, using Theorem 1, we show that the RS scheme achieves the
optimum DMT in the setup of single-antenna two-hop multiple-relay
networks where there exists no direct link neither between  the transmitter
and the receiver, nor between the relay nodes.

\begin{thm}
Assume a single-antenna half-duplex parallel relay scenario with $K$
non-interfering relays. The proposed SM scheme with $L=BK$,
$S=BK+1$, the path sequence $$\mathrm Q \equiv (\mathrm q_1, \dots,
\mathrm q_K, \mathrm q_1, \dots, \mathrm q_K, \dots, \mathrm q_1,
\dots, \mathrm q_K)$$ where $\mathrm q_k\equiv(0, k, K+1)$ and the
timing sequence $s_{i,j}=i+j-1$ achieves the diversity gain
\begin{equation}
d_{RS,NI}(r)=\max \left\{0,
K\left(1-r\right)- \frac{r}{B} \right\}, \label{eq:t1}
\end{equation}
which achieves the optimum DMT curve $d_{opt}(r)=K(1-r)^+$ as $B \to \infty$.
\end{thm}
\begin{proof}
First, according to the cut-set bound theorem \cite{cover_book}, the
point-to-point capacity of the uplink channel (the channel from the
transmitter to the relays) is an upper-bound on the achievable rate
of the network. Accordingly, the diversity-multiplexing curve of a
$1 \times K$ SIMO system which is a straight line (from the multiplexing
gain $1$ to the diversity gain $K$, i.e. $d_{opt}(r)=K(1-r)^+$) is
an upper-bound on the DMT of the network. Now, we prove that the
proposed RS scheme achieves the upper-bound on the DMT for
asymptotically large values of $S$.

As the relay pairs are non-interfering ($1 \leq k\leq K: \left\{k,
(k \mod K)+1 \right\} \notin E$), the result of Theorem 1 can be
applied. As a result
\begin{equation}
d_{RS, NI}(r) = \displaystyle \min_{\boldsymbol \mu \in \mathcal {\hat R}} \sum_{e \in E} \mu_e,
\end{equation}
where $\mathcal {\hat R} = \left\{  \boldsymbol \mu \left| \, \,
\mathbf 0 \leq \boldsymbol \mu \leq \mathbf 1, \displaystyle \sum_{i=1}^{BK}
\max_{1 \leq j \leq 2} \mu_{\left\{ \mathrm q_{(i-1) \mod K + 1}(j), \mathrm q_{(i-1) \mod K + 1}(j-1) \right\}}
\geq BK - (BK+1)r \right\} \right.$. Hence, we have
\begin{equation}
BK \left(1  - r - \frac{1}{BK}r\right)^+ \stackrel{(a)}{\leq} B
\sum_{k=1}^{K} \max \left\{ \mu_{\left\{ 0, k \right\}},
\mu_{\left\{ K+1, k \right\}} \right\} \leq B \displaystyle \sum_{e \in E} \mu_e,
\end{equation}
where $(a)$ results from the fact that every path $\mathrm q_k$ is
used $B$ times in the path sequence. Hence, DMT can be lower-bounded
as
\begin{equation}
d_{RS, NI}(r) \geq K \left(1  - r - \frac{1}{BK}r\right)^+. \label{eq:t11_1}
\end{equation}
On the other hand, considering the vector $\boldsymbol {\hat
\mu}=[\hat \mu_e]_{e \in E}$ where $\forall 1 \leq k \leq K: \hat
\mu_{\left\{0,k\right\}} = \left( 1 - r - \frac{1}{BK}r \right)^+$
and $\forall k,k' \neq 0: \hat \mu_{\left\{k,k'\right\}}=0$, it is
easy to verify that $\boldsymbol {\hat \mu} \in \mathcal {\hat R}$.
Hence,
\begin{equation}
d_{RS,NI}(r) \leq \sum_{e \in E} \hat \mu_e = K \left(1  - r - \frac{1}{BK}r\right)^+. \label{eq:t11_2}
\end{equation}
Combining \eqref{eq:t11_1} and \eqref{eq:t11_2} completes the proof.
\end{proof}

\textit{Remark 3-} Note that as long as the complement\footnote{For every undirected graph $G=(V,E)$, the complement of $G$ is a graph $H$ on the same vertices such that two vertices of $H$ are adjacent if and only if they are non-adjacent in $G$. \cite{graph_book}} of the induced sub-graph of $G$
on the relay nodes $\left\{1, 2, \dots, K \right\}$ includes a
Hamiltonian cycle \footnote{A Hamiltonian cycle is a simple cycle
$(v_1,v_2,\cdots ,v_K ,v_1)$ that goes exactly one time through each
vertex of the graph\cite{graph_book}.}, the result of Theorem 2
remains valid. However, the paths $\mathrm q_1, \mathrm q_2, \dots,
\mathrm q_K$ should be permuted in the path sequence according to
their orderings in the corresponding Hamiltonian cycle.

\begin{figure}[hbt]
  \centering
  \includegraphics[scale=0.8]{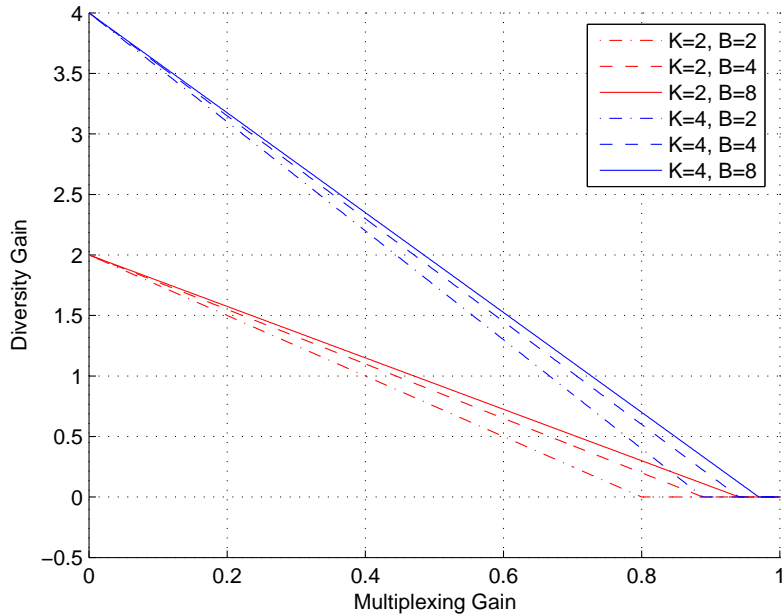}
\caption{DMT of RS scheme in parallel relay network for both ``interfering''
and ``non-interfering'' relaying scenarios and for different values of $K,B$.} \label{fig:dm_wi}
\end{figure}

According to (\ref{eq:t1}), we observe that the RS scheme achieves
the maximum multiplexing gain $1-\frac{1}{BK+1}$ and the maximum
diversity gain $K$, respectively, for the setup of non-interfering
relays. Hence, it achieves the maximum diversity gain for any finite
value of $B$. Also, knowing that no signal is sent to the receiver
in the first slot, the RS scheme achieves the maximum possible
multiplexing gain. Figure (\ref{fig:dm_wi}) shows the DMT of the
scheme for the case of non-interfering relays and various values of
$K$ and $B$.

\subsection{General Case}

In this section, we study the performance of the RS scheme in
general single-antenna multi-hop wireless networks and derive a
lower bound on the corresponding DMT. First, we show that the RS
scheme with the parameters defined in Theorem 2 achieves the optimum
DMT for the single-antenna parallel-relay networks when there is no
direct link between the transmitter and the receiver. Then, we
generalize the statement and provide a lower-bound on the DMT of the
RS scheme for the more general case.

As stated in the section ``System Model'', throughout the two-hop
network analysis, we slightly modify our notations to simplify the
derivations. Specifically, the output vector at the transmitter, the
input and the output vectors at the $k$'th relay, and the input
vector at the receiver are denoted as $\mathbf x$, $\mathbf r_k$,
$\mathbf t_k$ and $\mathbf y$, respectively. $h_k$ and $g_k$
represent the channel gain between the transmitter and the $k$'th
relay and the channel gain between the $k$'th relay and the
destination, respectively. $(k)$ and $(b)$ are defined as $(k)\equiv
\left((k-2) \mod K\right) +1$ and $(b)\equiv b - \lfloor
\frac{(k)}{K} \rfloor$. Finally, $i_{(k)}$, $\mathbf n_k$, $\mathbf
z$, and $\alpha_k$ denote the channel gain between the $k$'th and
the $(k)$'th relay nodes, the noise at the $k$'th relay and at the
receiver, and the amplification coefficient at the $k$'th relay.

\begin{figure}[t]
  \centering
  \includegraphics[scale=1.0]{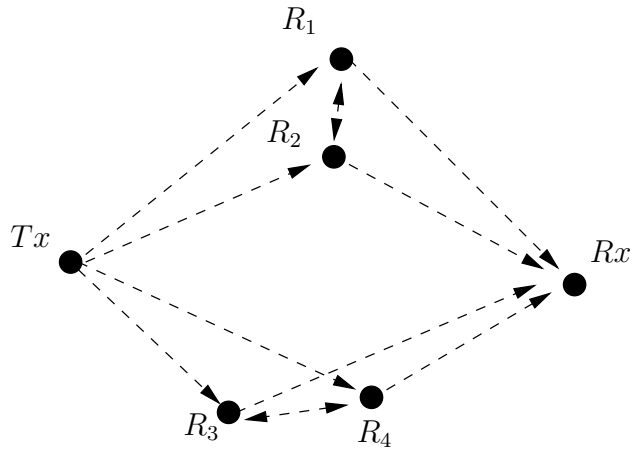}
\caption{An example of the half-duplex parallel relay network setup, relay nodes $\{1,2\}$ are disconnected from relay nodes $\{3,4\}$.}
\label{fig:model}
\end{figure}

Figure (\ref{fig:model}) shows a realization of this setup with $4$ relays. As observed, the relay set $\{1,2\}$ is disconnected from the relay set
$\{3,4\}$. In general, the output signal of any relay node $k'$ such that $\{k, k'\} \in E$ can interfere on the received signal of relay node $k$. However, in Theorem 3, the RS scheme is applied with the same parameters as in Theorem 2. Hence, when the transmitter is sending signal to the $k$'th relay in a time-slot, just the $(k)$'th relay is simultaneously transmitting and interferes at the $k$'th relay side. As an example, for the scenario shown in figure (\ref{fig:model}), we have
\begin{eqnarray}
\mathbf r_1  & = & h_1 \mathbf x + i_{4} \mathbf t_4 + \mathbf n_1, \nonumber \\
\mathbf r_2  & = & h_2 \mathbf x + \mathbf n_2. \nonumber
\end{eqnarray}

However, for the sake of simplicity, in the proof of the following theorem, we assume that all the relays interfere with each other. Hence, at the $k$'th relay, we have
\begin{equation}
\mathbf{r}_k = h_{k} \mathbf{x} + i_{(k)} \mathbf t_{(k)} + \mathbf n_k.
\end{equation}
According to the output power constraint, the amplification
coefficient is bounded as $\alpha _k \leq \sqrt {\frac{P}{P \left(
\left| h_k \right|^2 + \left| i_{(k)} \right|^2 \right) + 1}}$.
However, according to the signal boosting constraint imposed on the RS scheme, we have $|\alpha_k| \leq 1$. Hence, the amplification coefficient is equal to
\begin{equation}
\alpha_k = \min
\left\{ 1, \sqrt {\frac{P}{P \left( \left| h_k \right|^2 + \left| i_{(k)}
\right|^2 \right) + 1} } \right\}. \label{eq:alpha_constraint}
\end{equation}
In this manner, it is guaranteed that the noise terms of the
different relays are not boosted throughout the network. This is
achieved at the cost of working with the output power less than $P$.
On the other hand, we know that almost surely \footnote{By almost
surely, we mean its probability is greater than $1-P^{-\delta}$, for
any value of $\delta > 0$.} $\left| h_k \right|^2 , \left| i_{(k)}
\right|^2 \dot{\leq} 1$. Hence, almost surely, we have $\alpha _k
\doteq 1$. This point will be elaborated further in the proof of the
theorem. Now, we prove the DMT optimality of the RS scheme for general
single-antenna parallel-relay networks.

\begin{thm} \label{thm:DMT-n-ir}
Consider a single-antenna half-duplex parallel relay network with
$K>1$ interfering relays where there is no direct link between the
transmitter and the receiver. The diversity gain of the RS scheme
with the parameters defined in Theorem 2 is lower-bounded as
\begin{equation}
d_{RS,I}(r) \geq \max \left\{ 0, K \left( 1 - r \right) -
\frac{r}{B} \right\}. \label{eq:t2}
\end{equation}
Furthermore, the RS scheme achieves the optimum DMT
$d_{opt}(r)=K(1-r)^+$ as $B \to \infty$.
\end{thm}
\begin{proof}
First, we show that the entire channel matrix is equivalent to a
lower triangular matrix. Let us define $\mathbf{x}_{b,k},
\mathbf{n}_{b,k}, \mathbf{r}_{b,k}, \mathbf{t}_{b,k},
\mathbf{z}_{b,k}, \mathbf{y}_{b,k}$ as the portion of signals that
is sent or received in the $k$'th slot of the $b$'th sub-block. At
the receiver side, we have
\begin{eqnarray}
\mathbf{y}_{b,k} & = & g_{(k)} \mathbf{t}_{b, k} + \mathbf{z}_{b, k}
\nonumber \\
& = & g_{(k)} \alpha _{(k)} \left( \sum_{\begin{subarray}{c}1 \leq b_1\leq b, 1 \leq k_1 \leq K \\ b_1 K + k_1
< b K + k \end{subarray}}{p_{b-b_1, k, k_1}\left( h_{k_1}\mathbf{x}_{b_1,
k_1} + \mathbf{n}_{b_1, k_1}\right) }  \right) + \mathbf{z}_{b, k}.
\end{eqnarray}
Here, $p_{b, k, k_1}$ has the following recursive formula $p_{0, k,
k}=1, p_{b, k, k_1}=i_{((k))}\alpha_{((k))}p_{(b), (k), k_1}$. Defining
the square $BK \times BK$ matrices $\mathbf{G}= \mathbf{I}_B
\otimes \textit{diag}\left\{ g_1, g_2, \cdots ,g_K \right\}$,
$\mathbf{H}= \mathbf{I}_B \otimes \textit{diag}\left\{ h_1, h_2,
\cdots ,h_K \right\}$, $\mathbf{\Omega} = \mathbf{I}_B \otimes
\textit{diag}\left\{ \alpha _1, \alpha _2, \cdots ,\alpha _K
\right\}$, and
\begin{equation}
\mathbf{F}= \left(
\begin{array}{ccccc}
1 & 0 & 0 & 0 & \ldots \\
p_{0,2,1} & 1 & 0 & 0 & \ldots \\
p_{0,3,1} & p_{0, 3, 2} & 1 & 0 & \ldots \\
\vdots & \vdots & \vdots & \vdots & \ddots \\
p_{B-1, K, 1} & p_{B-1, K, 2} & \ldots & p_{0, K, K-1} & 1
\end{array}
 \right),
\end{equation}
where $\otimes$ is the Kronecker product \cite{matrix_book} of
matrices and $\mathbf{I}_B$ is the $B \times B$ identity matrix, and
the $BK \times 1$ vectors $\mathbf{x}\left(s\right)=[x_{1,1}(s),
x_{1, 2}(s), \cdots ,x_{B, K}(s)]^T$,
$\mathbf{n}\left(s\right)=\left[n_{1,1}\left(s\right), n_{1, 2}(s),
\cdots ,n_{B, K}(s)\right]^T$,
$\mathbf{z}\left(s\right)=[z_{1,2}(s), z_{1, 3}(s), \cdots ,z_{B+1,
1}(s)]^T$, and $\mathbf{y}\left(s\right)=[y_{1,2}(s), y_{1, 3}(s),
\cdots ,y_{B+1, 1}(s)]^T$, we have
\begin{equation}
\mathbf{y}\left(s\right) = \mathbf{G} \mathbf{\Omega} \mathbf{F}
\left( \mathbf{H} \mathbf{x}\left(s\right) +
\mathbf{n}\left(s\right) \right) + \mathbf{z}\left(s\right).
\label{eq:ref1}
\end{equation}
Here, we observe that the matrix of the entire channel is equivalent
to a lower triangular matrix of size $BK \times BK$ for a MIMO
system with a colored noise. The probability of outage of such a
channel for the multiplexing gain $r$ ($r \leq 1$) is defined as
\begin{equation}
\mathbb{P} \left\{ \mathcal{E} \right\}=\mathbb{P} \left\{ \log
 \left|\mathbf{I}_{BK} + P \mathbf{H}_{T}\mathbf{H}_{T}^{H}\mathbf{P}_n^{-1} \right| \leq
(BK+1)r \log\left( P \right) \right\},
\end{equation}
where $\mathbf{P}_n=\mathbf{I}_{BK}+\mathbf{G} \mathbf{\Omega}
\mathbf{F} \mathbf{F}^H \mathbf{\Omega}^H \mathbf{G}^H$, and
$\mathbf{H}_T=\mathbf{G} \mathbf{\Omega} \mathbf{F} \mathbf{H}$.
Assume $|h_k|^2=P^{-\mu_k}$, $|g_k|^2=P^{-\nu_k}$,
$|i_k|^2=P^{-\omega_k}$, and $\mathcal{R}$ as the region in
$\mathbb{R}^{3K}$ that defines the outage event $\mathcal{E}$ in
terms of the vector $[\boldsymbol \mu^T, \boldsymbol \nu^T,
\boldsymbol \omega^T]^T$, where $\boldsymbol \mu=\left[ \mu_1 \mu_2
\cdots \mu_K \right]^T, \boldsymbol \nu=\left[ \nu_1 \nu_2 \cdots
\nu_K \right]^T,\boldsymbol \omega=\left[ \omega_1 \omega_2 \cdots
\omega_K \right]^T$. The probability distribution function (and also
the complement of the cumulative distribution function) decays
exponentially as $P^{-P^{-\delta}}$ for positive values of $\delta$.
Hence, the outage region $\mathcal R$ is almost surely equal to
$\mathcal{R}_{+}=\mathcal{R} \bigcap \mathbb{R}_{+}^{3K}$. Now, we
have
\begin{eqnarray}
\mathbb{P} \left\{ \mathcal{E} \right\} & \stackrel{(a)}{\leq} &
\mathbb{P} \left\{ \left| \mathbf{H}_T \right|^2 \left| \mathbf{P}_n
\right|^{-1} \leq
P^{-BK \left( 1-r \right) +r}\right\} \nonumber \\
& \stackrel{(b)}{\leq} & \mathbb{P} \left\{ -B
\sum_{k=1}^{K}{\left(\mu_k+\nu_k- \min \left\{ 0, \mu_k, \omega_{(k)}
\right\}\right)} - \frac{BK\log(3) + \log \left| \mathbf{P}_{n} \right|}{\log
\left( P \right)} \leq -BK(1-r)+r
\right\} \nonumber \\
& \stackrel{(c)}{\dot{\leq}} & \mathbb{P} \left\{ - BK \frac{\log
\left[3 \left( B^2K^2+1 \right) \right]}{\log (P)} + BK\left( 1-r
\right) - r \leq B
\sum_{k=1}^{K}{(\mu_k + \nu_k)}, \mu_k,\nu_k,\omega_k \geq 0 \right\}. \label{eq:R_hat_wi}
\end{eqnarray}
Here, $(a)$ follows from the fact that for a positive semidefinite
matrix $\mathbf A$, we have $\left| \mathbf{I} + \mathbf{A} \right|
\geq \left| \mathbf{A} \right|$ and $(b)$ follows from the fact that
\begin{equation}
|\alpha_k|^2= \min \left\{ 1, \frac{P}{P^{1-\mu_k} + P^{1-\omega_{(k)}
 }+1} \right\} \geq \frac{1}{3} \min \left\{ 1,  P, P^{\mu_k
}, P^{\omega_{(k)}} \right\}  \nonumber
\end{equation}
and assuming $P$ is large enough such that $P \geq 1$. Finally, $(c)$ is proved as follows:

As $|\alpha_k| \leq 1$, we conclude $p_{n,k,k_1} \leq 1$. Hence, the sum of the entries of each row in $\mathbf{F}\mathbf{F}^H$ is less than $B^2K^2$. Now, consider the  matrix $\mathbf{A} \triangleq B^2 K^2 \mathbf I -\mathbf{F}\mathbf{F}^H$. From the above discussion, it follows that for every $i$, we have $A_{i,i} \geq \sum_{i \neq j}|A_{i,j}|$. Hence, for every vector $\mathbf x$, we have $\mathbf x^T \mathbf A \mathbf x \geq \sum_{i < j} |A_{i,j}| x_i^2 + |A_{i,j}| x_j^2 \pm 2|A_{i,j}|x_ix_j = \sum_{i < j} |A_{i,j}| \left(x_i \pm x_j \right)^2 \geq 0$, and as a result $\mathbf A$ is positive semidefinite, which implies that $\mathbf{F}\mathbf{F}^H \preccurlyeq B^2K^2 \mathbf{I}_{BK}$. Consequently, we have $\mathbf{P}_n \preccurlyeq \mathbf I_{BK} + B^2K^2 \mathbf{G} \mathbf{\Omega} \mathbf{\Omega}^H \mathbf{G}^H$. Moreover, Knowing the fact that $\mathbb P \left\{ \mathcal{R} \right\} \doteq \mathbb P \left\{
\mathcal{R}_{+} \right\}$, and conditioned on $\mathcal{R}_{+}$, one has $|g_k|^2 \leq 1$, which implies that $\mathbf {GG}^H \preccurlyeq \mathbf I$. Combining this with the fact that $\mathbf{\Omega} \mathbf{\Omega}^H \preccurlyeq \mathbf I$ (as $|\alpha_k|^2 \leq 1$, $\forall k$) yields  $\mathbf{P}_n \preccurlyeq \mathbf I_{BK} + B^2K^2 \mathbf{G} \mathbf{\Omega}
 \mathbf{\Omega}^H \mathbf{G}^H \preccurlyeq \left(B^2K^2 + 1 \right) \mathbf I_{BK}$. Moreover, conditioned on  $\mathcal{R}_{+}$, we have $\min \left\{ 0, \mu_k, \omega_{(k)} \right\} = 0$. This completes the proof of $(c)$.

On the other hand, for vectors $\boldsymbol{\mu}^0, \boldsymbol
{\nu}^0, \boldsymbol{\omega}^0 \geq \mathbf 0$, we have $\mathbb{P}
\left\{\boldsymbol{\mu} \geq \boldsymbol{\mu}^0, \boldsymbol{\nu}
\geq \boldsymbol{\nu}^0, \boldsymbol{\omega} \geq
\boldsymbol{\omega}^0 \right\} \doteq P^{-\mathbf{1} \cdot \left(
\boldsymbol{\mu}^0 + \boldsymbol{\nu}^0 + \boldsymbol{\omega}^0
\right)}$. Similar to the proof of Theorem 1, by taking derivative
with respect to $\boldsymbol \mu, \boldsymbol \nu$, we have
$f_{\boldsymbol \mu, \boldsymbol \nu}(\boldsymbol \mu, \boldsymbol
\nu) \doteq P^{- \mathbf 1 \cdot \left( \boldsymbol \mu +
\boldsymbol \nu \right)}$. Defining $l_0 \triangleq
- \frac{\log \left[3 \left( B^2K^2+1 \right) \right]}{\log (P)} +
\left( 1-r \right) - \frac{r}{BK} $, $\hat{\mathcal{R}} \triangleq  \left\{
\boldsymbol{\mu},\boldsymbol{\nu} \geq \mathbf{0}, \frac{1}{K}
\mathbf{1} \cdot \left( \boldsymbol{\mu} + \boldsymbol{\nu}\right)
\geq l_0 \right\}$, the cube $\mathcal I$ as $\mathcal I \triangleq  \left[0,
Kl_0 \right]^{2K}$, and for $1 \leq i \leq 2K$, $\mathcal{I}_i^c \triangleq [0,
\infty )^{i-1} \times [Kl_0, \infty ) \times [0, \infty )^{2K-i}$,
we observe
\begin{eqnarray}
\mathbb{P} \left\{ \mathcal E \right\} & \stackrel{(a)}{\dot{\leq}}
& \mathbb{P} \{ \hat{\mathcal R} \} \nonumber \\
& \stackrel{(b)}{\leq} & \int_{\mathcal{\hat{R}} \bigcap \mathcal{I}}{f_{\boldsymbol
\mu, \boldsymbol \nu}\left(\boldsymbol \mu, \boldsymbol \nu \right) d\boldsymbol \mu
d \boldsymbol \nu} + \sum_{i=1}^{2K}{\mathbb{P} \left\{ [\boldsymbol \mu^T,
\boldsymbol \nu^T]^T \in  \mathcal{\hat{R}} \cap \mathcal{I}_i^c \right\}}
\nonumber
\\
&\dot{\leq} & \mbox{vol} (\mathcal{\hat{R}} \cap \mathcal{I})
P^{\displaystyle -\min_{\left[ \boldsymbol{\mu}_0^T, \boldsymbol{\nu}_0^T \right]^T \in
\mathcal{\hat{R}} \bigcap \mathcal{I}} \mathbf{1} \cdot \left(
\boldsymbol{\mu}_0 + \boldsymbol{\nu}_0 \right) } + 2K P^{-Kl_0}
\nonumber \\
& \stackrel{(c)}{\doteq} & P^{-Kl_0} \nonumber \\
& \doteq & P^{-\left[K \left( 1 - r \right) - \frac{r}{B} \right]}.
\label{eq:t2_r_wi}
\end{eqnarray}
Here, $(a)$ follows from (\ref{eq:R_hat_wi}), $(b)$ results from writing $\mathcal{\hat{R}}$ as $\left(\mathcal{\hat{R}} \bigcap \mathcal I \right) \bigcup \left[ \bigcup_{i=1}^M \left( \mathcal{\hat{R}} \bigcap \mathcal I_i^c\right) \right] $ and using the union bound on the probability, and $(c)$ follows from the
fact that $\mathcal{\hat{R}} \bigcap \mathcal{I}$ is a bounded
region whose volume is independent of $P$. (\ref{eq:t2_r_wi})
completes the proof of Theorem 3.
\end{proof}

\textit{Remark 4-} The argument in Theorem 3 is valid no matter what
the induced graph of $G$ on the relay nodes is. More precisely, the
DMT of the RS scheme can be lower-bounded as \eqref{eq:t2} as long
as $\left\{0, K+1 \right\} \notin E$ and $\left\{0, k\right\},
\left\{K+1, k\right\} \in E$. One special case is that the complement of the induced
subgraph of $G$ on the relay nodes includes a Hamiltonian cycle
which is analyzed in Theorem 2. Here, we observe that the
lower-bound on DMT derived in \eqref{eq:t2} is tight as shown in
Theorem 2.

Figure (\ref{fig:dm_wi}) shows the DMT of the RS scheme for varying
number of $K$ and $B$. Noting the proof of Theorem 3, we can easily
generalize the result of Theorem 3 and provide a lower-bound on the
DMT of the RS scheme for general single-antenna multi-hop
multiple-relay networks.

\begin{thm}
Consider a half-duplex single-antenna multiple-relay network with
the connectivity graph $G=(V, E)$ operated under the RS scheme with
$L$ paths, $S$ slots, and the path sequence $\left(\mathrm p_1,
\mathrm p_2, \dots, \mathrm p_L\right)$. Defining $\beta_e$ for
each $e \in E$ as the number of paths in the path sequence that go
through $e$, then the DMT of the RS scheme is lower-bounded as
\begin{equation}
d_{RS}(r) \geq \frac{L}{\displaystyle \max_{e \in E} \beta_e} \left(1 - \frac{S}{L}r \right)^+. \label{eq:t4_lb}
\end{equation}
\end{thm}
\begin{proof}
First, similar to the proof of Theorem 3, we show that the entire
channel matrix is lower triangular. At the receiver side, we have
\begin{equation}
 \mathbf y_{K+1,i}  =  \prod_{j=1}^{l_i} h_{\left\{\mathrm p_i(j), \mathrm p_i(j-1)\right\}} \alpha_{i,j} \mathbf x_{0,i} +
\sum_{j < i}  f_{i, j} \mathbf x_{0,j} + \sum_{j \leq i, m \leq l_j}  q_{i, j, m} \mathbf n_{j, m}.\label{eq:t4_r_side}
\end{equation}
Here, $\mathbf x_{0,i}$ is the vector transmitted at the transmitter
side during the $s_{i,1}$'th slot as the input for the $i$'th path,
$\mathbf y_{K+1,i}$ is the vector received at the receiver side
during the $s_{i,l_i}$'th slot as the output for $i$'th path,
$f_{i,j}$ is the interference coefficient which relates the input of
the $j$'th path ($j < i$) to the output of the $i$'th path, $\mathbf
n_{j,m}$ is the noise vector during the $s_{j,m}$'th slot at the
$\mathrm p_j(m)$'th node, and finally, $q_{i, k, m}$ is the
coefficient which relates $\mathbf n_{k,m}$ to $\mathbf y_{K+1,i}$.
Note that as the timing sequence satisfies the noncausal
interference assumption, the summation terms in \eqref{eq:t4_r_side}
do not exceed $i$. Moreover, for the sake of brevity, we define
$\alpha_{i,l_i}=1$. Defining $\mathbf x(s) = \left[ x_{0,1}
\left(s\right)  x_{0,2} \left(s\right) \cdots x_{0,L}\left(s\right)
\right]^T$, $\mathbf y(s) = \left[ y_{K+1,1}\left(s\right)
y_{K+1,2}\left(s\right) \cdots y_{K+1,L}\left(s\right) \right]^T$,
and $\mathbf n(s) = \left[ n_{1,1}\left(s\right)
n_{1,2}\left(s\right) \cdots n_{L,l_L}\left(s\right) \right]^T$, we
have the following equivalent lower-triangular matrix between the
end nodes:
\begin{equation}
\mathbf y(s) = \mathbf H_T \mathbf x(s) + \mathbf Q \mathbf n(s).
\end{equation}
Here,
\begin{equation}
\mathbf H_T= \left(
\begin{array}{cccc}
f_{1,1} &  0 &  0 & \ldots \\
f_{2,1} & f_{2,2} &  0  & \ldots \\
\vdots & \vdots & \vdots & \ddots \\
 f_{L,1} &  f_{L,2} & \ldots &  f_{L,L}
\end{array}
 \right),
\end{equation}
where $f_{i,i}=\displaystyle \prod_{j=1}^{l_i} h_{\left\{\mathrm p_i(j), \mathrm p_i(j-1)\right\}} \alpha_{i,j}$, and
\begin{equation}
\mathbf Q= \left(
\begin{array}{ccccccc}
 q_{1,1,1} &  \ldots &  q_{1,1,l_1} &  0 &  0 &  0 & \ldots \\
 q_{2,1,1} &  \ldots &  q_{2,1,l_1} & \ldots    &  q_{2,2,l_2} &  0 & \ldots \\
\vdots & \vdots & \vdots & \vdots & \vdots & \vdots & \ddots \\
 q_{L,1,1} &  q_{L,1,2} & \ldots & \ldots& \ldots&  q_{L,L,l_L-1} &  q_{L,L,l_L}
\end{array}
 \right).
\end{equation}
Let us define $\mu_e$ for every $e \in E$  such that
$|h_e|^2=P^{-\mu_e}$. First, we observe that similar to the proof of
 Theorem \ref{thm:DMT-n-ir}, it can be shown that i)~$\alpha_{i,j}
\doteq 1$ with probability 1\footnote{More precisely, with
probability greater than $1-P^{-\delta}$, for any $\delta > 0$.}, ii)~we can restrict ourselves to the region $\mathbb R_+$, i.e., the region $\boldsymbol{\mu} > \mathbf 0$. These two facts imply that $|q_{i, j, m}| \dot \leq 1$. This  
means there exists a constant $c$ which depends just on
the topology of the graph $G$ and the path sequence such that
$\mathbf P_n \triangleq \mathbf Q \mathbf Q^H \preccurlyeq c \mathbf
I_L$ (by a similar argument as in the proof of Theorem 3). Hence, similar to the arguments in the equation series
\eqref{eq:R_hat_wi}, the outage probability can be bounded as
\begin{eqnarray}
\mathbb P \left\{ \mathcal E \right\} & = & \mathbb P \left\{ \left|
\mathbf I_{L} + P \mathbf H_T \mathbf H_T^H \mathbf P_n^{-1} \right| \leq P^{Sr} \right\}\nonumber \\
& \dot \leq & \mathbb P \left\{ \left| \mathbf H_T \right| \left| \mathbf H_T^H \right| \leq P^{Sr-L} \right\} \nonumber \\
& = & \mathbb P \left\{ \sum_{e \in E} \beta_e \mu_e \geq L - Sr \right\} \nonumber \\
& \doteq & \mathbb P \left\{\boldsymbol \mu \geq \mathbf 0, \sum_{e \in E} \beta_e \mu_e \geq (L - Sr)^+ \right\},
\end{eqnarray}
where $\beta_e$ is the number of paths in the path sequence that
pass through $e$. Knowing that $\mathbb P \left\{ \boldsymbol \mu \geq
\boldsymbol \mu^0 \right\} \doteq P^{-\mathbf 1 \cdot \boldsymbol
\mu}$ and computing the derivative, we have $f_{\boldsymbol
\mu}(\boldsymbol \mu) = P^{-\mathbf 1 \cdot \boldsymbol \mu}$.
Defining $\mathcal R = \left\{ \boldsymbol \mu >
\mathbf 0, \sum_{e \in E} \beta_e \mu_e \geq (L - Sr)^+ \right\}$
and applying the results of equation series \eqref{eq:t2_r_wi}, we
obtain
\begin{eqnarray}
\mathbb P \left\{ \mathcal E \right\} & \dot \leq &  P^{- \displaystyle \min_{\boldsymbol \mu \in \mathcal R}
\mathbf 1 \cdot \boldsymbol \mu } \stackrel{(a)}{=} P^{\displaystyle  - \frac{L}{\max_{e \in E} \beta_e}
\left(1 - \frac{S}{L}r \right)^+}, \label{eq:t4_last_eq}
\end{eqnarray}
where $(a)$ follows from the fact that for every $\boldsymbol \mu
\in \mathcal R$, $\left(L - Sr \right)^+ \leq \sum_{e \in E} \beta_e
\mu_e \leq \max_{e \in E} \beta_e \sum_{e \in E} \mu_e$ which implies that $\sum_{e \in E} \mu_e = \mathbf 1 \cdot \boldsymbol{\mu} \geq \frac{\left(L - Sr \right)^+}{\max_{e \in E} \beta_e}$, and on the
other hand, defining $\boldsymbol \mu^\star$ such that $\boldsymbol
\mu^\star(\hat e) = \frac{(L-Sr)^+}{\beta_{\hat e}}$ where $\hat e =
\underset{e \in E}{\mbox{argmax}}~\beta_e$ and otherwise
$\boldsymbol \mu^\star(e) = 0$, we have $\boldsymbol \mu^\star \in
\mathcal R$ and $\mathbf 1 \cdot \boldsymbol \mu^\star =
\frac{L}{\max_{e \in E} \beta_e} \left(1 - \frac{S}{L}r \right)^+$.
\eqref{eq:t4_last_eq} completes the proof of Theorem 4.
\end{proof}

\textit{Remark 5-} The lower-bound of \eqref{eq:t1_lb} can also be proved by using the lower-bound of \eqref{eq:t4_lb} obtained for DMT of the general RS scheme. In order to prove this, one needs to apply the RS scheme with the same path sequence and timing sequence used in the proof of \eqref{eq:t1_lb} in Theorem 1. Putting $S=L_0 d_G$ and $S\leq l_G L$ in \eqref{eq:t4_lb} and noting that for all $e \in E$, we have $\beta_e \in \left\{0, L_0 \right\}$, \eqref{eq:t1_lb} is easily obtained.

\textit{Remark 6-} It should be noted that \eqref{eq:t1_ub} is yet an upper-bound for the DMT of the RS scheme, i.e., even for the case of interfering relays. This is due to the fact that in the proof of \eqref{eq:t1_ub} the non-interfering relaying assumption is not used. However, by employing the RS scheme with causal-interfering relaying and applying \eqref{eq:t4_lb}, one can find a bigger family of graph topologies that can achieve \eqref{eq:t1_ub}. Such an example is the two-hop relay network studied in Theorem 3. Another example is the case that $G$ is a directed acyclic graph (DAG)\footnote{A directed acyclic graph $G$ is a directed graph that has no directed cycles.} and the relays are operating in the full-duplex mode. Here, the argument is similar to that of \textit{Remark 2}. Assume that each  $\mathrm { \hat p_i}$ is used $L_0$ times in the path sequence in the form that $\mathrm p_{(i-1)L_0+j} \triangleq \mathrm { \hat p_i}, 1 \leq j \leq L_0$. Let us modify the timing sequence as $s_{i, j}=i + j - 1 + \displaystyle \sum_{k=1}^{\lceil \frac{i}{L_0} \rceil - 1} \hat l_k$ which results in $S=L+\sum_{i=1}^{d_G} l_i$. Here, it is easy to verify that only non-causal interference exists between the signals corresponding to different paths.  However, by considering the paths in the reverse order or equivalently reversing the time axis, the paths can be observed with the causal interference. Hence, the result of Theorem 4 is still valid for such paths. Here, knowing that for all $e \in E$, we have $\beta_e \in \left\{0, L_0 \right\}$ and applying \eqref{eq:t4_lb}, we have $d_{RS}(r) \geq d_G \left( 1 - r - \frac{\sum_{i=1}^{d_G} l_i}{L_0d_G} \right)^+$ which achieves \eqref{eq:t1_ub} for asymptotically large values of $L_0$. This fact is also observed by \cite{vkumar}.

\subsection{Multiple-Access Multiple-Relay Scenario}
In this subsection, we generalize the result of Theorem 3 to the
multiple-access scenario aided by multiple relay nodes. Here,
similar to Theorem 3, we assume that there is no direct link between
each transmitter and the receiver. However, no restriction is
imposed on the induced subgraph of $G$ on the relay nodes. Assuming
having $M$ transmitters, we show that for the rate sequence $r_1
\log (P), r_2 \log (P), \dots, r_M\log (P)$, in the asymptotic case
of $B \to \infty$ ($B$ is the number of sub-blocks), the RS scheme achieves the diversity gain $d_{SM,
MAC}(r_1, r_2, \dots, r_M)=K \left(1 - \sum_{m=1}^{M}{r_m}
\right)^{+}$, which is shown to be optimum due to the cut-set bound
on the cutset between the relays and the receiver. Here, the
notations are slightly modified compared to the ones used in Theorem
3 to emphasize the fact that multiple signals are transmitted from
multiple transmitters. Throughout this subsection and the next one,
$\mathbf{x}_m$ and $h_{m, k}$ denote the transmitted vector at the
$m$'th transmitter and the Rayleigh channel coefficient between the
$m$'th transmitter and the $k$'th relay, respectively. Hence, at the
received side of the $k$'th relay, we have
\begin{equation}
\mathbf{r}_k = \sum_{m=1}^{M}{h_{m, k} \mathbf{x}_m} + i_{(k)}
\mathbf{t}_{(k)} + \mathbf{n}_k,
\end{equation}
where $\mathbf{x}_m$ is the transmitted vector of the $m$'th
sender. The amplification coefficient at the $k$'th relay is set to
\begin{equation}
\alpha_k = \min \left\{ 1, \sqrt { \frac{P}{P \left( \sum_{m=1}^{M}{\left|
h_{m,k} \right|^2} + \left| i_{(k)} \right|^2 \right) + 1}} \right\}.
\end{equation}
Here, the RS scheme is applied with the same path sequence and
timing sequence as in the case of Theorem 2 and 3. However, it
should be mentioned that in the current case, during the slots that
the transmitter is supposed to transmit the signal, i.e. in the
$s_{i,1}$'th slot, all the transmitters send their signals
coherently. Moreover, at the receiver side, after receiving the $BK$
vectors corresponding to the outputs of the $BK$ paths, the
destination node decodes the messages $\omega_1, \omega_2, \dots,
\omega_K$ by joint-typical decoding of the received vectors in the
corresponding $BK$ slots and the transmitted signal of all the
transmitters, i.e., in the same way that joint-typical decoding works
in the multiple access setup~\cite{cover_book}. Now, we prove the
main result of this subsection.

\begin{thm}
Consider a multiple-access channel consisting of $M$ transmitting
nodes aided by $K>1$ half-duplex relays. Assume there is no direct
link between the transmitters and the receiver. The RS scheme with
the path sequence and timing sequence defined in Theorems 2 and 3
achieves a diversity gain of
\begin{equation}
d_{RS,MAC}(r_1, r_2, \dots, r_M) \geq \left[ K \left( 1 -
\sum_{m=1}^{M}{r_m} \right) - \frac{\sum_{m=1}^{M}{r_m}}{B}
\right]^+, \label{eq:t3}
\end{equation}
where $r_1, r_2, \dots, r_M$ are the multiplexing gains corresponding to users
$1,2,\dots,M$. Moreover, as $B \to \infty$, it achieves the optimum
DMT which is $d_{opt, MAC}(r_1, r_2, \dots, r_M)=K \left(1 -
\sum_{m=1}^{M}{r_m} \right)^{+}$.
\end{thm}
\begin{proof}
At the receiver side, we have
\begin{eqnarray}
\mathbf{y}_{b,k} & = & g_{(k)} \mathbf{t}_{b, k} + \mathbf{z}_{b, k}
\nonumber \\
& = & g_{(k)} \alpha _{(k)} \left( \sum_{\begin{subarray}{c} 1
\leq b_1 \leq b, 1 \leq k_1 \leq K \\ b_1 K + k_1 < b K + k \end{subarray}
}{p_{b-b_1, k, k_1}\left(
\sum_{m=1}^{M}{h_{m,k_1}\mathbf{x}_{m, b_1, k_1}} + \mathbf{n}_{b_1,
k_1}\right) }  \right) + \mathbf{z}_{b, k}, \nonumber \\
\end{eqnarray}
where $p_{b, k, k_1}$ is defined in the proof of Theorem 3 and
$\mathbf{x}_{m, b, k}$ represents the transmitted signal of the
$m$'th sender in the $k$'th slot of the $b$'th sub-block. Similar to
(\ref{eq:ref1}), we have
\begin{equation}
\mathbf{y}\left(s\right) = \mathbf{G} \mathbf{\Omega} \mathbf{F}
\left( \sum_{m=1}^{M}{ \mathbf{H}_m \mathbf{x}_m \left(s\right)} +
\mathbf{n}\left(s\right) \right) + \mathbf{z}\left(s\right),
\end{equation}
where  $\mathbf{H}_m= \mathbf{I}_B \otimes \textit{diag}\left\{
h_{m,1}, h_{m,2}, \cdots ,h_{m,K} \right\}$,
$\mathbf{x}_m\left(s\right)=[x_{m,1,1}(s), x_{m,1, 2}(s), \cdots
,x_{m, B, K}(s)]^T$, and $\mathbf {y}_s , \mathbf {n}_s, \mathbf
{z}_s, \mathbf G , \mathbf \Omega , \mathbf F$ are defined in the
proof of Theorem 3. Similarly, we observe that the entire channel
from each of the transmitters to the receiver acts as a MIMO channel
with a lower triangular matrix of size $BK \times BK$.

Here, the outage event occurs whenever there exists a subset
$\mathcal S \subseteq \left\{ 1, 2, \dots, M \right\}$ of the
transmitters such that
\begin{equation}
I \left( \mathbf{x}_{\mathcal S}(s) ; \mathbf y(s) |
\mathbf{x}_{\mathcal S^c} (s) \right) \leq \left( BK + 1 \right)
\left( \sum_{m \in \mathcal S}{r_m} \right) \log (P).
\end{equation}
This event is equivalent to
\begin{equation}
\log \left|\mathbf{I}_{BK} + P
\mathbf{H}_{T}\mathbf{H}_{T}^{H}\mathbf{P}_n^{-1} \right| \leq
(BK+1) \left( \sum_{m \in \mathcal S}{r_m} \right) \log\left( P
\right).
\end{equation}
where $\mathbf{P}_n$ is defined in the proof of Theorem 3, $\mathbf{H}_T =
\mathbf G \mathbf \Omega \mathbf F \mathbf{H}_{\mathcal S}$, and
\begin{equation}
\mathbf{H}_{\mathcal S}= \mathbf{I}_B \otimes \textit{diag}\left\{
\sqrt{\sum_{m \in \mathcal S}\left| h_{m,1}\right|^2}, \sqrt{\sum_{m
\in \mathcal S}\left| h_{m,2}\right|^2}, \cdots , \sqrt{\sum_{m \in
\mathcal S}\left| h_{m,K}\right|^2} \right\}.
\end{equation}
Defining such an event as $\mathcal E_{\mathcal S}$ and the outage
event as $\mathcal E$, we have
\begin{eqnarray}
\mathbb {P} \left\{ \mathcal E \right\} & = & \mathbb {P} \left\{
\bigcup_{\mathcal S \subseteq \left\{1, 2, \dots, M \right\}}
\mathcal E_S \right\} \nonumber \\
& \leq & \sum_{\mathcal S \subseteq \left\{1, 2, \dots, M
\right\}}{\mathbb {P} \left\{ \mathcal E_S \right\}} \nonumber \\
& \leq & (2^M - 1)\max_{\mathcal S \subseteq \left\{1, 2, \dots, M
\right\}}{\mathbb {P} \left\{ \mathcal E_S \right\}} \nonumber \\
& \doteq & \max_{\mathcal S \subseteq \left\{1, 2, \dots, M
\right\}}{\mathbb {P} \left\{ \mathcal E_S \right\}}.
\label{eq:t3_r00}
\end{eqnarray}
Hence, it is sufficient to upper-bound $\mathbb {P} \left\{ \mathcal
E_S \right\}$ for all $\mathcal S$.

Defining $\hat{\mathbf{H}}_{\mathcal S} = \mathbf{I}_B \otimes
\textit{diag}\left\{ \max_{m \in \mathcal S}\left| h_{m,1} \right|,
\max_{m \in \mathcal S}\left| h_{m,2} \right|, \cdots , \max_{m \in
\mathcal S}\left| h_{m,K} \right| \right\}$, we have
$\hat{\mathbf{H}}_{\mathcal S} \hat{\mathbf{H}}_{\mathcal S}^H
\preccurlyeq \mathbf{H}_{\mathcal S} \mathbf{H}_{\mathcal S}^H$.
Therefore,
\begin{eqnarray}
\mathbb {P} \left\{ \mathcal E_S \right\} & \leq & \mathbb P \left
\{ \log \left|\mathbf{I}_{BK} + P \mathbf G \mathbf \Omega \mathbf F
\hat{\mathbf{H}}_{\mathcal S}\hat{\mathbf{H}}_{\mathcal S}^{H}
\mathbf F^H \mathbf \Omega^H \mathbf G^H \mathbf{P}_n^{-1} \right|
\leq (BK+1) \left( \sum_{m \in \mathcal S}{r_m} \right) \log\left( P
\right) \right\} \nonumber \\ & \triangleq & \mathbb P \left\{
\hat{\mathcal E}_{\mathcal S} \right\}. \label{eq:t3_r01}
\end{eqnarray}

Assume $\max_{m \in \mathcal S}|h_{m, k}|^2=P^{-\mu_k}$, and
$|g_k|^2=P^{-\nu_k}$, $|i_k|^2=P^{-\omega_k}$, and $\mathcal{R}$ as
the region in $\mathbb{R}^{3K}$ that defines the outage event $\hat
{\mathcal{E}_{\mathcal S}}$ in terms of the vector $[\boldsymbol
\mu^T, \boldsymbol \nu^T, \boldsymbol \omega^T]^T$. Similar to the proof of Theorem 3, we have $\mathbb{P} \left\{ \mathcal R \right\} \doteq
\mathbb{P} \left\{ \mathcal{R}_{+} \right\}$ where $\mathcal{R}_{+}
= \mathcal R \bigcap \mathbb{R}_{+}^{3K}$. Rewriting the equation
series of (\ref{eq:R_hat_wi}), we have
\begin{eqnarray}
\mathbb{P} \left\{ \hat{\mathcal E}_{\mathcal S} \right\} &
\dot{\leq} & \mathbb{P} \left\{ - BK \frac{\log \left[ 3 \left(
B^2K^2+1 \right) \right]}{\log (P)} + BK\left( 1- \sum_{m \in
\mathcal S}{r_m} \right) - \sum_{m \in \mathcal S}{r_m} \leq B
\sum_{k=1}^{K}{(\mu_k + \nu_k)},\right. \nonumber \\
&&  \mu_k,\nu_k,\omega_k \geq 0 \Bigg\}. \label{eq:t3_r1}
\end{eqnarray}

On the other hand, as $\left\{h_{m,k}\right\}$'s are independent
random variables, we conclude that for $\boldsymbol{\mu}^0,
\boldsymbol{\nu}^0 \geq \mathbf 0$, we have
$\mathbb{P} \left\{\boldsymbol{\mu} \geq \boldsymbol{\mu}^0,
\boldsymbol{\nu} \geq \boldsymbol{\nu}^0 \right\} \doteq P^{-\mathbf{1} \cdot \left(
\left| \mathcal S \right| \boldsymbol{\mu}^0 + \boldsymbol{\nu}^0  \right)}$. Similar to the proof of Theorem 3,
by computing the derivative with respect to $\boldsymbol \mu,
\boldsymbol \nu$, we have $f_{\boldsymbol \mu, \boldsymbol
\nu}(\boldsymbol \mu, \boldsymbol \nu) \doteq P^{- \mathbf 1 \cdot
\left( \left| \mathcal S \right| \boldsymbol \mu + \boldsymbol \nu
\right)}$. Defining  $l_0 \triangleq  - \frac{\log
\left[ 3 \left(B^2K^2+1 \right) \right] }{\log (P)} + \left( 1-
\sum_{m \in \mathcal S}{r_m} \right) - \frac{\sum_{m \in \mathcal
S}{r_m}}{BK} $, the region $\hat{\mathcal{R}}$ as
$\hat{\mathcal{R}} \triangleq \left\{ \boldsymbol \mu,\boldsymbol \nu \geq
\mathbf{0}, \frac{1}{K} \mathbf{1} \cdot \left( \boldsymbol \mu +
\boldsymbol \nu\right) \geq l_0 \right\}$, the cube $\mathcal I$ as
$\mathcal I \triangleq \left[0, Kl_0 \right]^{2K}$, and for $1 \leq i \leq
2K$, $\mathcal{I}_i^c=[0, \infty )^{i-1} \times [Kl_0, \infty )
\times [0, \infty )^{2K-i}$, we have
\begin{eqnarray}
\mathbb{P} \left\{ \hat{\mathcal E}_{\mathcal S} \right\} &
\stackrel{(a)}{\dot{\leq}}
& \mathbb{P} \{ \hat{\mathcal R} \} \nonumber \\
& \leq & \int_{\mathcal{\hat{R}} \bigcap \mathcal{I}}{f_{\boldsymbol
\mu, \boldsymbol \nu}\left(\boldsymbol \mu, \boldsymbol \nu \right) d\boldsymbol \mu
d \boldsymbol \nu} + \sum_{i=1}^{2K}{\mathbb{P} \left\{ [\boldsymbol \mu^T,
\boldsymbol \nu^T]^T \in  \mathcal{\hat{R}} \cap \mathcal{I}_i^c \right\}}
\nonumber
\\
&\dot{\leq} & \mbox{vol} (\mathcal{\hat{R}} \cap \mathcal{I})
P^{\displaystyle -\min_{\left[ \boldsymbol{\mu}, \boldsymbol{\nu} \right] \in
\mathcal{\hat{R}} \bigcap \mathcal{I}} \mathbf{1} \cdot \left(
\left| \mathcal S \right| \boldsymbol{\mu} + \boldsymbol{\nu} \right) }
+ 2K P^{-Kl_0}
\nonumber \\
& \stackrel{(b)}{\doteq} & P^{-Kl_0} \nonumber \\
& \doteq & P^{-\left[K \left( 1 - \sum_{m \in \mathcal S}{r_m}
\right) - \frac{\sum_{m \in \mathcal S}{r_m}}{B} \right]}.
\label{eq:t3_r2}
\end{eqnarray}
Here, (a) follows from (\ref{eq:t3_r1}) and (b) follows from the
fact that $\mathcal{\hat{R}} \bigcap \mathcal{I}$ is a bounded
region whose volume is independent of $P$ and the fact that $\min_{\left[ \boldsymbol{\mu}, \boldsymbol{\nu} \right] \in
\mathcal{\hat{R}} \bigcap \mathcal{I}} \mathbf{1} \cdot \left(
\left| \mathcal S \right| \boldsymbol{\mu} + \boldsymbol{\nu} \right) = K l_0 $, which is achieved by having $\boldsymbol{\mu}=\mathbf 0$. Comparing
(\ref{eq:t3_r00}), (\ref{eq:t3_r01}) and (\ref{eq:t3_r2}), we
observe
\begin{equation}
\mathbb {P} \left\{ \mathcal E \right\} \dot{\leq} \max_{\mathcal S
\subseteq \left\{1, 2, \dots, M \right\}}{\mathbb {P} \left\{
\mathcal E_S \right\}} \dot{\leq} \max_{\mathcal S \subseteq
\left\{1, 2, \dots, M \right\}}{\mathbb{P} \left\{ \hat{\mathcal
E}_{\mathcal S} \right\}} \dot{\leq} P^{-\left[K \left( 1 -
\sum_{m=1}^{M}{r_m} \right) - \frac{\sum_{m=1}^{M}{r_m}}{B}
\right]}.
\end{equation}
Next, we prove that $K \left( 1 - \sum_{m=1}^{M}{r_m} \right)^+$ is
an upper-bound on the diversity gain of the system corresponding to
the sequence of rates $r_1, r_2, \dots, r_M$. We have
\begin{equation}
\mathbb {P} \left\{ \mathcal E \right\} \geq \mathbb {P} \left\{
\max_{p\left(\mathbf t_1, \mathbf t_2, \dots , \mathbf t_K
\right)}{I \left(\mathbf t_1, \mathbf t_2, \dots , \mathbf t_K ;
\mathbf y \right)} \leq  \left( \sum_{m=1}^{M}{r_m} \right) \log (P)
\right\} \stackrel{(a)}{\doteq} P^{-K\left( 1 - \sum_{m=1}^{M}{r_m}
\right)^+}.
\end{equation}
Here, (a) follows from the DMT of the point-to-point MISO channel
proved in \cite{ zheng_tse}. This completes the proof.
\end{proof}

\textit{Remark 7-} The argument of Theorem 5 is valid for the general
case in which any arbitrary set of relay pairs are non-interfering.

\textit{Remark 8-} In the \textit{symmetric} situation for which the
multiplexing gains of all the users are equal to say $r$, the
lower-bound in \eqref{eq:t3} takes a simple form. First, we observe
that the maximum multiplexing gain which is simultaneously
achievable by all the users is $\frac{1}{M} \cdot \frac{BK}{BK+1}$.
Noting that no signal is sent to the receiver in $\frac{1}{BK+1}$
portion of the time, we observe that the RS scheme achieves the
maximum possible symmetric multiplexing gain for all the users.
Moreover, from (\ref{eq:t3}), we observe that the RS scheme achieves
the maximum diversity gain of $K$ for any finite value of $B$, which
turns out to be tight as well. Finally, the lower-bound on the DMT
of the RS scheme is simplified to $\left[ K \left( 1 - Mr\right) -
\frac{Mr}{B} \right]^+$ for the \textit{symmetric} situation.

\subsection{Multiple-Access Single Relay Scenario}
As we observe, the arguments of Theorems 2, 3 and 5 concerning DMT
optimality of the RS scheme are valid for the scenario of having
multiple relays ($K > 1$). Indeed, for the single relay scenario,
the RS scheme is reduced to the simple amplify-and-forward relaying
in which the relay listens to the transmitter in the first half of
the frame and transmits the amplified version of the received signal
in the second half. However, like the case of non-interfering relays
studied in \cite{yang_belfiore2}, the DMT optimality arguments are
no longer valid. On the other hand, we show that the DDF scheme
achieves the optimum DMT for this scenario.

\begin{thm}
Consider a multiple-access channel consisting of $M$ transmitting
nodes aided by a single half-duplex relay. Assume that all the
network nodes are equipped with a single antenna and there is no
direct link between the transmitters and the receiver. The
amplify-and-forward scheme achieves the following DMT
\begin{equation}
d_{AF,MAC}(r_1, r_2, \dots , r_M) = \left( 1 - 2 \sum_{m=1}^{M}{r_m}
\right)^{+}. \label{eq:mac_1r_af}
\end{equation}
However, the optimum DMT of the network is
\begin{equation}
d_{MAC}(r_1, r_2, \dots , r_M) = \left( 1 -
\frac{\sum_{m=1}^{M}{r_m}}{1-\sum_{m=1}^{M}{r_m}}
 \right)^{+}, \label{eq:mac_1r_ddf}
\end{equation}
which is achievable by the DDF scheme of \cite{azarian}.
\end{thm}
\begin{proof}
First, we show that the DMT of the AF scheme follows
(\ref{eq:mac_1r_af}). At the receiver side, we have
\begin{equation}
\mathbf y = g \alpha \left( \sum_{m=1}^{M}{h_m \mathbf x_m} +
\mathbf n \right) + \mathbf z,
\end{equation}
where $h_m$ is the channel gain between the $m$'th transmitter and
the relay, $g$ is the down-link channel gain, and $\alpha = \sqrt{
\frac{P}{P \sum_{m=1}^{M}{\left| h_m \right|^2} + 1} }$ is the
amplification coefficient. Defining the outage event $\mathcal{E_S}$
for a set $\mathcal S \subseteq \{ 1, 2, \dots, M\}$, similar to the
case of Theorem 5, we have
\begin{eqnarray}
\mathbb P \left\{ \mathcal{E_S} \right\} & = & \mathbb P \left\{ I
\left( \mathbf{x}_{\mathcal S} ; \mathbf y | \mathbf{x}_{\mathcal
S^c} \right) < 2 \left( \sum_{m \in \mathcal S}{r_m} \right) \log
(P) \right\}
 \nonumber \\
& = & \mathbb P \left\{
 \log \left( 1 + P \left( \sum_{m \in \mathcal S}{\left| h_m \right|^2} \right) \left| g \right|^2
\left| \alpha \right|^2 \left( 1 +   \left| g \right|^2 \left|
\alpha \right|^2 \right)^{-1} \right) <  2 \left( \sum_{m \in
\mathcal S}{r_m} \right) \log (P) \right\} \nonumber \\
&\doteq & \mathbb{P}\left\{\left( \sum_{m \in \mathcal S}{\left| h_m
\right|^2} \right) |g|^2|{\alpha}|^2 \min \left\{ 1,
\frac{1}{|g|^2|{\alpha}|^2} \right\} \leq
P^{-\left(1-2\sum_{m \in \mathcal S}{r_m}\right)} \right\} \nonumber \\
& \stackrel{(a)}{\doteq} &  \mathbb{P} \left\{\sum_{m \in \mathcal S}{\left| h_m
\right|^2} \leq  P^{-\left(1-2\sum_{m \in \mathcal S}{r_m}\right)}
\right\} + \nonumber \\
& & \mathbb{P} \left\{   \left( \sum_{m \in \mathcal S}{\left|
h_m \right|^2} \right) |g|^2 |\alpha|^2 \leq
P^{-\left(1-2\sum_{m \in \mathcal S}{r_m}\right)} \right\}  \nonumber \\
& \stackrel{(b)}{\doteq} &  \mathbb{P} \left\{\sum_{m \in \mathcal S}{\left| h_m
\right|^2} \leq  P^{-\left(1-2\sum_{m \in \mathcal S}{r_m}\right)}
\right\} + \nonumber \\
& & \mathbb{P} \left\{ |g|^2  \left( \sum_{m \in \mathcal S}{\left|
h_m \right|^2} \right) \min \left\{P , \frac{1}{
\sum_{m=1}^{M}{\left| h_m \right|^2} }\right\} \leq
P^{-\left(1-2\sum_{m \in \mathcal S}{r_m}\right)} \right\}  \nonumber \\
& \stackrel{(a)}{\doteq} & \mathbb{P} \left\{\sum_{m \in \mathcal S}{\left| h_m
\right|^2} \leq  P^{-\left(1-2\sum_{m \in \mathcal S}{r_m}\right)}
\right\}
 + \nonumber \\
 & & \mathbb P \left\{ \left| g \right|^2 \left( \sum_{m \in \mathcal S}{ \left| h_m \right|^2 } \right) \leq P^{-2 \left( 1 - \sum_{m
\in \mathcal S}{r_m} \right)} \right\} + \nonumber \\
& & \mathbb P \left\{ \frac {\left| g
\right|^2 \sum_{m \in \mathcal S}{ \left| h_m \right|^2 } }{\sum_{m=1}^{M}{\left| h_{m} \right|^2}} \leq P^{-
\left(1-2\sum_{m \in \mathcal S}{r_m}\right) } \right\} \label{eq:1r_1}.
\end{eqnarray}
In the above equation, $(a)$ comes from the fact that $\mathbb P \{\min (X,Y) \leq z\} = \mathbb P \left\{ (X \leq z) \bigcup (Y \leq z)\right\} \doteq \mathbb P \{X \leq z\} + \mathbb P \{Y \leq z \}$. $(b)$ follows from the fact that $|\alpha|^2$ can be asymptotically written as $\min \left\{P , \frac{1}{
\sum_{m=1}^{M}{\left| h_m \right|^2} }\right\}$. 

Since $\{|h_m|^2\}_{m=1}^M$ are i.i.d. random variables with exponential distribution, it follows that $\sum_{m \in \mathcal S}{\left| h_m
\right|^2}$ has Chi-square distribution with $2 |\mathcal S|$ degrees of freedom, which implies that 
\begin{eqnarray}
 \mathbb{P} \left\{\sum_{m \in \mathcal S}{\left| h_m
\right|^2} \leq  P^{-\left(1-2\sum_{m \in \mathcal S}{r_m}\right)}
\right\} \doteq P^{-|\mathcal S|\left(1-2\sum_{m \in \mathcal S}{r_m}\right)}. \label{eq:1}
\end{eqnarray}
To compute the second term in (\ref{eq:1r_1}), defining $\epsilon_1 \triangleq P^{-2 \left(1-\sum_{m \in \mathcal S}{r_m}\right)}$, we have
\begin{eqnarray}
\mathbb P \left \{ \left| g \right|^2 \left( \sum_{m \in \mathcal S}{ \left| h_m \right|^2 } \right) \leq \epsilon_1 \right\}  &\stackrel{(a)}{\dot \geq}&  \mathbb P \left \{ \left| g \right|^2  \leq \epsilon_1 \right\}   \notag\\
&\doteq& \epsilon_1, \label{eq:t4_t1}
\end{eqnarray}
where $(a)$ follows from the fact that as $\sum_{m \in \mathcal S}{\left| h_m
\right|^2}$ has Chi-square distribution, we have $\sum_{m \in \mathcal S}{\left| h_m
\right|^2} \dot \leq 1$ with probability one (more
precisely, with a probability greater than $1 - P^{-\delta}$ for
every $\delta > 0$).
On the other hand, we have
\begin{eqnarray}
 \mathbb P \left \{ \left| g \right|^2 \left( \sum_{m \in \mathcal S}{ \left| h_m \right|^2 } \right) \leq \epsilon_1\right\}
 &\leq& \mathbb P \left\{ \left| g  \right|^2 |h_m|^2\leq \epsilon_1 \right\} \notag\\
 &\doteq& \epsilon_1. \label{eq:t4_t2}
\end{eqnarray}
Putting (\ref{eq:t4_t1}) and (\ref{eq:t4_t2}) together, we have
\begin{equation}
 \mathbb P \left \{ \left| g \right|^2 \left( \sum_{m \in \mathcal S}{ \left| h_m \right|^2 } \right) \leq \epsilon_1 \right\} \doteq \epsilon_1 . \label{eq:t4_t3}
\end{equation}
Now, to compute the third term in (\ref{eq:1r_1}), defining $\epsilon_2 \triangleq P^{- \left(1-2\sum_{m \in \mathcal S}{r_m}\right)}$, we observe
\begin{equation}
 \epsilon_2 \doteq \mathbb P \left\{ \left| g \right|^2 \leq \epsilon_2 \right\}
 \leq \mathbb P \left\{ \left| g \right|^2 \frac {\sum_{m \in \mathcal S}{ \left| h_m \right|^2 }}
 {\sum_{m=1}^{M}{\left| h_m \right|^2 }} \leq \epsilon_2 \right\} \stackrel{(a)}{\dot{\leq}}
 \mathbb P \left\{  \left| g \right|^2  \left( \sum_{m \in \mathcal S}{ \left| h_m \right|^2 } \right)
 \leq \epsilon_2 \right\} \stackrel{(b)}{\doteq} \epsilon_2 . \nonumber
\end{equation}
Here, $(a)$ follows from the fact that with probability one, we have $\sum_{m=1}^{M}{\left| h_m \right|^2 } \dot{\leq} 1$ and $(b)$ follows
from (\ref{eq:t4_t3}). As a result
\begin{equation}
 \mathbb P \left\{ \left| g \right|^2 \frac {\sum_{m \in \mathcal S}{ \left| h_m \right|^2 }}
 {\sum_{m=1}^{M}{\left| h_m \right|^2 }} \leq \epsilon_2 \right\} \doteq \epsilon_2 \label{eq:t4_t4}
\end{equation}
From (\ref{eq:1}), (\ref{eq:t4_t3}), and (\ref{eq:t4_t4}), we have
\begin{equation}
\mathbb P \left\{ \mathcal{E_S} \right\} \doteq P^{- \left| \mathcal S \right|
\left( 1 - 2 \sum_{m \in \mathcal S}{r_m} \right)^+} + P^{-2 \left( 1 - \sum_{m \in \mathcal S}{r_m} \right)^+}
+ P^{- \left( 1 - 2 \sum_{m \in \mathcal S}{r_m} \right)^+} \doteq P^{-\left( 1 - 2 \sum_{m \in \mathcal S}{r_m} \right)^+}. \label{eq:t4_t5}
\end{equation}
Observing (\ref{eq:t4_t5}) and applying the argument of (\ref{eq:t3_r00}), we have
\begin{equation}
\mathbb P \left\{ \mathcal E \right\} \doteq \max_{\mathcal S \subseteq
\left\{1,2,\dots,M \right\}}{\mathbb P \left\{ \mathcal{E_S} \right\} } \doteq P^{-\left( 1 - 2 \sum_{m=1}^{M}{r_m} \right)^+}.
\end{equation}
This completes the proof for the AF scheme. Now, to compute the DMT
of the DDF scheme, let us assume that the relay listens to the
transmitted signal for the $l$ portion of the time until it can decode it perfectly. Hence, we have
\begin{equation}
 l = \min \left\{ 1, \max_{\mathcal S \subseteq \left\{ 1, 2, \dots, M \right\} }
 {\frac { \left( \sum_{m \in \mathcal S}{r_m} \right) \log (P) }
 {\log \left( 1+ \left( \sum_{m \in \mathcal S}{ \left| h_m \right|^2 } \right) P \right) } } \right\}.
\end{equation}
The outage event occurs whenever the relay can not transmit the
re-encoded information in the remaining portion of the time. Hence,
we have
\begin{equation}
\mathbb P \left\{ \mathcal E \right\} \doteq \mathbb P \left\{ \left( 1- l \right)
\log \left( 1 + \left| g \right|^2 P \right) < \left( \sum_{m=1}^{M}{r_m} \right) \log (P) \right\}.
\end{equation}
Assuming $|h_m|^2=P^{-\mu_m}$ and $|g|^2=P^{-\nu}$, at high SNR, we
have
\begin{equation}
l \approx \min \left\{1, \max_{\mathcal S \subseteq \left\{1,2, \dots , M \right\} }
\frac {\sum_{m \in \mathcal S}{r_m}}{1-\min_{m \in \mathcal S}{\mu_m} } \right\} .
\end{equation}
Equivalently, an outage event occurs whenever
\begin{equation}
\left( 1 - \max_{\mathcal S \subseteq \left\{1,2, \dots , M \right\} }
\frac {\sum_{m \in \mathcal S}{r_m}}{1-\min_{m \in \mathcal S}{\mu_m} } \right)
\left( 1- \nu \right) < \sum_{m=1}^{M}{r_m}. \label{eq:t4_out_reg}
\end{equation}
In order to find the probability of the outage event, we first find an upper-bound on the outage probability and then, we show that this upper-bound is indeed tight. Defining $R=\sum_{m=1}^{M}{r_M}$ and $\mu =
\sum_{m=1}^{M}{\mu_m}$, we have
\begin{equation}
R  \stackrel{(a)}{>}  \left( 1 - \frac {\sum_{m \in \mathcal S_0}{r_m}}
{1-\min_{m \in \mathcal S_0}{\mu_m} } \right) \left( 1 - \nu \right) >
\left( 1 - \frac {R}{1- \mu } \right) \left( 1 - \nu \right). \label{eq:t4_t6}
\end{equation}
Here, $(a)$ follows from (\ref{eq:t4_out_reg}). Equivalently,
\begin{equation}
R \stackrel{(a)}{>} \frac{(1 - \mu)(1 - \nu)}{(1 - \mu) + (1 - \nu)}
> \frac{1 - \mu - \nu}{(1 - \mu) + (1 - \nu)}, \label{eq:t4_t7}
\end{equation}
where $(a)$ follows from (\ref{eq:t4_t6}). It can be easily checked
that (\ref{eq:t4_t7}) is equivalent to
\begin{equation}
R > ( 1 - R ) ( 1 - \mu - \nu). \label{eq:t4_t8}
\end{equation}
In other words, any vector point $\left[\mu_1, \mu_2, \dots,
\mu_M, \nu \right]$ in the outage region $\mathcal R$, i.e., the
region that satisfies (\ref{eq:t4_out_reg}), also satisfies (\ref{eq:t4_t8}). As a result, defining $\mathcal{R}'$ as the region defined by (\ref{eq:t4_t8}), we have
\begin{eqnarray}
 \mathbb P \{\mathcal E\} &\leq& \mathbb P \{\boldsymbol{\pi} \in \mathcal {R}'\},
\end{eqnarray}
 where $\boldsymbol{\pi} \triangleq \left[\mu_1, \mu_2, \dots,
\mu_M, \nu \right]$. Similar to the approach used in the proofs of Theorems 3 and 5, $\mathbb P \{\boldsymbol{\pi} \in \mathcal {R}'\}$ can be computed as
\begin{eqnarray}
 \mathbb P \{\boldsymbol{\pi} \in \mathcal {R}'\} \doteq P^{-\frac{R}{1-R}}.
\end{eqnarray}
Hence,
\begin{eqnarray} \label{ttttl}
 \mathbb P \{ \mathcal E\} \dot \leq P^{-\frac{R}{1-R}}.
\end{eqnarray}
For lower-bounding the outage probability, we note that all the vectors $[\mu_1, \cdots, \mu_M, \nu]$ for which $\mu_m > 0, m=1, \cdots, M$ and $\nu > \frac{R}{1-R}$, lie in the outage region defined in (\ref{eq:t4_out_reg}). In other words,
\begin{eqnarray} \label{ttttu}
 \mathbb P \{ \mathcal E\} &\geq& \mathbb P \left \{\boldsymbol{\pi} > \left[0, \cdots,0, \frac{R}{1-R} \right] \right \} \notag\\
&\doteq& P ^{-\frac{R}{1-R}}.
\end{eqnarray}
Combining (\ref{ttttl}) and (\ref{ttttu}) yields
\begin{eqnarray} \label{tttt}
 \mathbb P \{ \mathcal E\} &\doteq& P^{-\frac{R}{1-R}} \notag\\
&=& P^{-\frac{\sum_{m=1}^M r_m}{1- \sum_{m=1}^M r_m}},
\end{eqnarray}
which completes the proof for the DMT analysis of the DDF scheme.

Next, we prove that the DDF scheme achieves the optimum DMT. As the
channel from the transmitters to the receiver is a degraded version
of the channel between the transmitters and the relay, similar to
the argument of \cite{cover} for the case of single-source
single-relay, we can easily show that the decode-forward
strategy achieves the capacity of the network for each realization
of the channels. Now, consider the realization in which for all $m$ we
have, $\left |h_m \right|^2 \leq \frac{1}{M}$. As we know, $\mathbb
P \left\{\forall m: \left |h_m \right|^2 \leq \frac{1}{M} \right\}
\doteq 1$. Let us assume in  the optimum decode-and-forward strategy, the relay
spends $l$ portion of the time for listening  to the transmitter. According to the Fano's inequality \cite{cover_book}, to make the probability of error in decoding the transmitters' message
at the relay side approach zero, we should have $l \log \left(1 +
\frac{P}{l} \sum_{m=1}^M \left|h_m\right|^2 \right) \geq \left(
\sum_{m=1}^M r_m \right) \log (P)$. Accordingly, we should have
$l \geq \sum_{m=1}^M r_m$. On the other hand, in order that the
receiver can decode the relay's message with a vanishing probability
of error in the remaining portion of the time, we should have
$\left( 1 - l \right) \log \left( 1 + \frac{P}{1-l} \left| g
\right|^2 \right) \geq \sum_{m=1}^M r_m \log (P)$. Hence, we have $\mathbb
P \left\{ \mathcal E \right\} \geq \mathbb P \left\{ \left| g
\right|^2 \leq cP^{-\left( 1 -
\frac{\sum_{m=1}^{M}{r_m}}{1-\sum_{m=1}^{M}{r_m}}
 \right)}, \forall m: \left |h_m \right|^2 \leq \frac{1}{M}  \right\} \doteq P^{-\left( 1 -
\frac{\sum_{m=1}^{M}{r_m}}{1-\sum_{m=1}^{M}{r_m}}
 \right)^+}$, for a constant $c$. This completes the proof.
\end{proof}

\begin{figure}[hbt]
  \centering
  \includegraphics[scale=0.7]{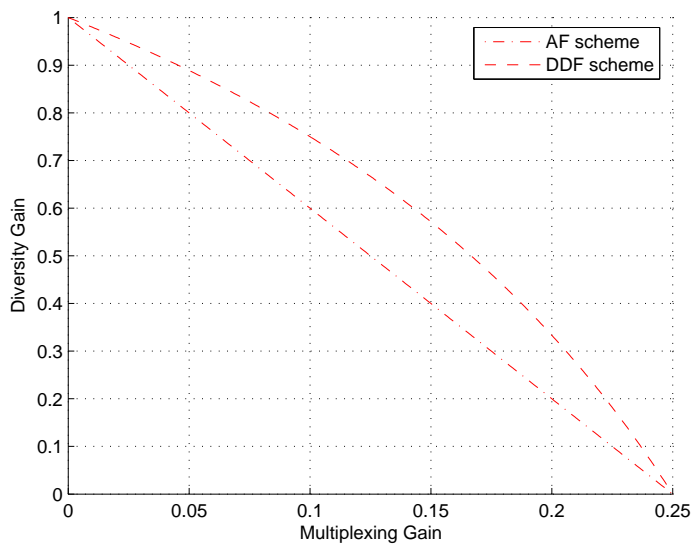}
\caption{Diversity-Multiplexing Tradeoff of AF scheme versus the
optimum and DDF scheme for multiple access single relay channel
consisting of $M=2$ transmitters assuming \textit{symmetric}
transmission, i.e. $r_1=r_2=r$.} \label{fig:dm_mac}
\end{figure}

Figure \ref{fig:dm_mac} shows DMT of the AF scheme and the DDF scheme for multiple access single relay setup consisting of $M=2$ transmitters assuming \textit{symmetric} situation, i.e. $r_1=r_2=r$. As can be observed in this figure, although the AF scheme achieves the maximum multiplexing gain and maximum diversity gain, it does not achieve the optimum DMT in any other points of the tradeoff region.

\section{Maximum Diversity Achievability Proof in General Multi-Hop Multiple-Antenna Scenario}
In this section, we consider our proposed RS scheme and prove that
it achieves the maximum diversity gain between two end-points in a
general multiple-antenna multi-hop network (no additional
constraints imposed). However, in this general scenario, it can not
achieve the optimum DMT. Indeed, we show that in order to achieve
the optimum DMT, in some scenarios, multiple interfering nodes have
to transmit together during the same slot.

\begin{thm}
Consider a relay network with the connectivity graph $G=(V,E)$ and
$K$ relays, in which  each two adjacent nodes are connected through a
Rayleigh-fading channel. Assume that all the network nodes are equipped with
multiple antennas. Then, by properly choosing the path sequence, the proposed RS scheme achieves the maximum
diversity gain of the network which is equal to
\begin{equation}
d_{G} = \min_{\mathcal S} w_G(\mathcal S),
\end{equation}
where $\mathcal S$ is a cut-set on $G$.
\end{thm}
\begin{proof}
First, we show that $d_G$ is indeed an upper-bound on the
diversity-gain of the network. To show this, we do not consider the
half-duplex nature of the relay nodes and assume that they operate
in full-duplex mode. Consider a cut-set $\mathcal S$ on $G$. We have
\begin{eqnarray} \label{mimo1}
\mathbb P \left\{ \mathcal E \right\} & \stackrel{(a)}{\dot \geq} & \mathbb P \left\{ I\left( X\left( \mathcal S \right);Y\left( \mathcal S^c \right) | X\left( \mathcal S^c \right) \right) < R \right\} \nonumber \\
& \stackrel{(b)}{=} & \mathbb P \left\{\sum_{k \in \mathcal S^c} I\left( X\left( \mathcal S \right);Y_k | Y\left( \mathcal S^c  / \left\{1,2, \dots, k \right\} \right),X\left( \mathcal S^c \right) \right) < R \right\} \nonumber \\
& \stackrel{(c)}{\geq} & \prod_{k \in \mathcal S^c} \mathbb P \left\{ I\left( X\left( \mathcal S \right);Y_k | X\left( \mathcal S^c \right) \right) < \frac{R}{\left| \mathcal S^c \right|} \right\} \nonumber \\
& \stackrel{(d)}{\doteq} & \prod_{k \in \mathcal S^c}P^{-\left| \left\{ e \in E | k \in e, e \cap \mathcal S \neq \oslash \right\} \right|} \nonumber \\
& \doteq & P^{-w_G\left( \mathcal S \right)},
\end{eqnarray}
where $R$ is the target rate which does not scale with $P$ (i.e., $r=0$). Here, $(a)$ follows from the cut-set bound theorem
\cite{cover_book} and the fact that for the rates above the
capacity, the error probability approaches one (according to Fano's
inequality \cite{cover_book}), $(b)$ follows from the chain rule on the  mutual
information \cite{cover_book}, $(c)$ follows from the
facts that i)~$\left(Y_k, X\left(\left\{0,1,\dots,K+1\right\}\right),
Y\left( \mathcal S^c  / \left\{1,2, \dots, k \right\}
\right)\right)$ form a Markov chain \cite{cover_book} and as a
result, $I\left( X\left( \mathcal S \right);Y_k | Y\left( \mathcal
S^c  / \left\{1,2, \dots, k \right\} \right),X\left( \mathcal S^c
\right) \right) \leq I\left( X\left( \mathcal S \right);Y_k |
X\left( \mathcal S^c \right) \right)$, and ii)~$I\left( X\left( \mathcal S \right);Y_k |
X\left( \mathcal S^c \right) \right)$ depends only on the channel matrices between $X\left( \mathcal S \right)$ and $Y_k$ and as all the channels in the network are independent of each other, it follows that the events $$\left\{ I\left( X\left( \mathcal S \right);Y_k | X\left( \mathcal S^c \right) \right) < \frac{R}{\left| \mathcal S^c \right|} \right\}_{k \in \mathcal S^c} $$ are mutually independent, and finally $(d)$ follows
from the diversity gain of the MISO channel. Considering all possible cut-sets on $G$ and using (\ref{mimo1}), we have
\begin{equation}
\mathbb P \left\{ \mathcal E \right\} \dot \geq P^{-\min_{\mathcal S}w_G \left( \mathcal S \right)}.
\end{equation}

Now, we prove that this bound is indeed achievable by the RS scheme.
First, we provide the path sequence needed to achieve the maximum
diversity gain. Consider the graph $\hat G = (V, E, w)$ with the
same set of vertices and edges as the graph $G$ and the weight
function $w$ on the edges as $w_{\left\{a,b \right\}}= N_a N_b$.
Consider the maximum-flow algorithm \cite{graph_book} on $\hat G$
from the source node $0$ to the sink node $K+1$. Since the weight
function is integer over the edges, according to the Ford-Fulkerson
Theorem \cite{graph_book}, one can achieve the maximum flow which is
equal to the minimum cut of $\hat G$ or $d_G$ by the union of
elements of a sequence $\left( \mathrm p_1, \mathrm p_2, \dots,
\mathrm p_{d_G}\right)$ of paths ($L = d_G$). We show that this family of paths
are sufficient to achieve the optimum diversity. Here, we do not
consider the problem of selecting the path timing sequence
$\left\{s_{i,j} \right\}$. We just assume that a timing sequence
$\left\{s_{i,j} \right\}$ with the 4 requirements defined in the
third section exists. However, it should be noted that as we
consider the maximum diversity throughout the theorem, we
are not concerned with $\frac{S}{L}$. Hence, we can select the path
timing sequence such that no two paths cause interference on each
other.

Noting that the received signal at each node is multiplied by a random isotropically distributed unitary matrix, at the receiver side we have
\begin{eqnarray}
 \mathbf y_{K+1,i} & = & \mathbf H_{K+1, \mathrm p_i(l_i -1)} \alpha_{i,l_i-1} \mathbf U_{i,l_i-1}
 \mathbf H_{\mathrm p_i(l_i -1), \mathrm p_i(l_i -2)} \alpha_{i,l_i-2} \mathbf U_{i,l_i-2}
 \cdots \alpha_{i,1} \mathbf U_{i,1} \mathbf H_{\mathrm p_i(1), 0} \mathbf x_{0,i} + \nonumber \\
& &\sum_{j < i} \mathbf X_{i, j} \mathbf x_{0,j} + \sum_{j \leq i, m
\leq l_j} \mathbf Q_{i, j, m} \mathbf n_{j,
m}.\label{eq:MIMO_div_model}
\end{eqnarray}
Here, $\mathbf x_{0,i}$ is the vector transmitted at the transmitter
side during the $s_{i,1}$'th slot as the input for the $i$'th path,
$\mathbf y_{K+1,i}$ is the vector received at the receiver side
during the $s_{i,l_i}$'th slot as the output for $i$'th path, $\mathbf U_{i,j}$ denotes the multiplied unitary matrix at the $\mathrm p_i (j)$'th node of the $i$th path,  
$\mathbf X_{i,j}$ is the interference matrix which relates the input
of the $j$'th path ($j < i$) to the output of the $i$'th path,
$\mathbf n_{j,m}$ is the noise vector during the $s_{j,m}$'th slot
at the $\mathrm p_j(m)$'th node of the network, and finally,
$\mathbf Q_{i, k, m}$ is the matrix which relates $\mathbf n_{k,m}$
to $\mathbf y_{K+1,i}$. Notice that as the timing sequence satisfies
the noncausal interference assumption, the summation terms in
\eqref{eq:MIMO_div_model} do not exceed $i$. Defining $\mathbf x(s)
= \left[ \mathbf x_{0,1}^T\left(s\right) \mathbf
x_{0,2}^T\left(s\right) \cdots \mathbf x_{0,L}^T\left(s\right)
\right]^T$, $\mathbf y(s) = \left[ \mathbf y_{K+1,1}^T\left(s\right)
\mathbf y_{K+1,2}^T\left(s\right) \cdots \mathbf
y_{K+1,L}^T\left(s\right) \right]^T$, and $\mathbf n(s) = \left[
\mathbf n_{1,1}^T\left(s\right) \mathbf n_{1,2}^T\left(s\right)
\cdots \mathbf n_{L,l_L}^T\left(s\right) \right]^T$, we have the
following equivalent block lower-triangular matrix between the end
nodes
\begin{equation} \label{mimo_vec}
\mathbf y(s) = \mathbf H_T \mathbf x(s) + \mathbf Q \mathbf n(s).
\end{equation}
Here,
\begin{equation}
\mathbf H_T= \left(
\begin{array}{cccc}
\mathbf X_{1,1} &  \mathbf 0 & \mathbf 0 & \ldots \\
\mathbf X_{2,1} & \mathbf X_{2,2} & \mathbf 0  & \ldots \\
\vdots & \vdots & \vdots & \ddots \\
\mathbf X_{L,1} & \mathbf X_{L,2} & \ldots & \mathbf X_{L,L}
\end{array}
 \right),
\end{equation}
where $\mathbf X_{i,i}=\mathbf H_{K+1, \mathrm p_i(l_i -1)} \alpha_{i,l_i-1} \mathbf U_{i,l_i-1}
 \mathbf H_{\mathrm p_i(l_i -1), \mathrm p_i(l_i -2)} \alpha_{i,l_i-2} \mathbf U_{i,l_i-2}
 \cdots \alpha_{i,1} \mathbf U_{i,1} \mathbf H_{\mathrm p_i(1), 0}$, and
\begin{equation}
\mathbf Q= \left(
\begin{array}{ccccccc}
\mathbf Q_{1,1,1} &  \ldots & \mathbf Q_{1,1,l_1} & \mathbf 0 & \mathbf 0 & \mathbf 0 & \ldots \\
\mathbf Q_{2,1,1} &  \ldots & \mathbf Q_{2,1,l_1} & \ldots    & \mathbf Q_{2,2,l_2} & \mathbf 0 & \ldots \\
\vdots & \vdots & \vdots & \vdots & \vdots & \vdots & \ddots \\
\mathbf Q_{L,1,1} & \mathbf Q_{L,1,2} & \ldots & \ldots& \ldots& \mathbf Q_{L,L,l_L-1} & \mathbf Q_{L,L,l_L}
\end{array}
 \right).
\end{equation}
Having (\ref{mimo_vec}), the outage probability can be written as
\begin{eqnarray}
\mathbb P \left\{ \mathcal E \right\} & =& \mathbb P \left\{ \left| \mathbf I_{L} + P \mathbf H_T \mathbf H_T^H \mathbf P_n^{-1} \right| < 2^{SR} \right\},
\end{eqnarray}
where $\mathbf P_n = \mathbf Q \mathbf Q^H$. First, similar to the
proof of theorem \ref{thm:DMT-n-ir}, we can show that $\alpha_{i,j}
\doteq 1$ with probability 1\footnote{More precisely, with
probability greater than $1-P^{-\delta}$ for any $\delta > 0$.}, and
also show that there exists a constant $c$ which depends just on the
topology of graph $G$ and the path sequence such that $\mathbf P_n
\preccurlyeq c \mathbf I_L$. Assume that for each $\{a,b\} \in E$,
$\lambda_{\max} \left( \mathbf H_{a,b} \right)=P^{-\mu_{\left\{a,b
\right\}}}$, where $\lambda_{\max} \left( \mathbf A \right)$ denotes
the greatest eigenvalue of $\mathbf A \mathbf A^H$. Also, assume that
\begin{eqnarray}
\gamma_{i,j} \triangleq \left| \mathbf v_{r,\max}^H\left( \mathbf H_{\left\{\mathrm p_i(j+1),
\mathrm p_i(j)\right\}} \right) \mathbf U_{i,j} \mathbf
v_{l,\max}\left( \mathbf H_{\left\{\mathrm p_i(j), \mathrm p_i(j-1)\right\}} \mathbf U_{i,j-1} \mathbf H_{\left\{\mathrm p_i(j-1), \mathrm p_i(j-2)\right\}} \dots \mathbf H_{\left\{\mathrm p_i(1), 0 \right\}} \right) \right|^2 = P^{-\nu_{i,j}},
\end{eqnarray}
 where $\mathbf
v_{l,\max}\left( \mathbf A \right)$ and $\mathbf v_{r,\max}\left(
\mathbf A \right)$ denote the left and the right eigenvectors of
$\mathbf A$ corresponding to $\lambda_{\max} \left( \mathbf A
\right)$, respectively.  The outage probability can be upper-bounded as
\begin{eqnarray}
\mathbb P \left\{ \mathcal E \right\} & \stackrel{(a)}{\leq} & \mathbb P \left\{ \lambda_{\max}\left(\left(\mathbf H_T \mathbf H_T^H \mathbf P_n^{-1} \right)^{\frac{1}{2}} \right) \leq \left( 2^{SR} -1 \right) P^{-1} \right\} \nonumber \\
& \stackrel{(b)}{\leq} & \mathbb P \left\{ \lambda_{\max}\left(\mathbf H_T \right) \leq c \left(  2^{SR} -1 \right) P^{-1} \right\} \nonumber \\
&  \stackrel{(c)}{\leq} &\mathbb P \left\{\bigcap_{i=1}^{L}\left( \lambda_{\max}\left(\mathbf X_{i,i} \right) \leq c \left(  2^{SR} -1 \right) P^{-1} \right) \right\} \nonumber \\
&  \stackrel{(d)}{\leq} & \mathbb P \left\{ \bigcap_{i=1}^{L} \left( \sum_{j=1}^{l_i}
\mu_{\left\{\mathrm p_i(j), \mathrm p_i(j-1)\right\}} +
\sum_{j=1}^{l_i-1} \nu_{i,j} \geq 1 - \log\frac{c \left(2^{SR} -1
\right)}{P} \right) \right\} \notag\\
&\stackrel{(e)}{\doteq}& \mathbb P \left\{ \bigcap_{i=1}^{L} \left( \sum_{j=1}^{l_i}
\mu_{\left\{\mathrm p_i(j), \mathrm p_i(j-1)\right\}} +
\sum_{j=1}^{l_i-1} \nu_{i,j} \geq 1  \right) \right\}.
\end{eqnarray}
In the above equation, $(a)$ follows from the fact that 
$ 1+ \lambda_{\max} (\mathbf A^{\frac{1}{2}}) \leq \left| \mathbf I + \mathbf A\right|,$
for a positive semi-definite matrix $\mathbf A$. $(b)$ results from $\mathbf P_n
\preccurlyeq c \mathbf I_L$. $(c)$ follows from the fact that $\lambda_{\max} (\mathbf H_T) \geq \max_i \lambda_{\max} (\mathbf X_{i,i})$. To obtain $(d)$, we first show that 
\begin{eqnarray} \label{mimo2}
\lambda_{\max} (\mathbf{AUB}) \geq \lambda_{\max} (\mathbf{A})\lambda_{\max} (\mathbf{A}) \left| \mathbf v_{r,\max}^H (\mathbf A) \mathbf U \mathbf v_{l, \max} (\mathbf B) \right|^2,
\end{eqnarray}
for any matrices $\mathbf A$, $\mathbf U$ and $\mathbf B$. To show this, we write
\begin{eqnarray}
\lambda_{\max} (\mathbf{AUB}) &=& \max_{\begin{subarray}{c}
\mathbf x\\
\|\mathbf x\|^2 =1
                                      \end{subarray}
} \left\| \mathbf x^H  \mathbf{AUB} \right\|^2 \notag\\
&\geq& \left\| \mathbf v_{l,\max} (\mathbf A) \mathbf{AUB}\right\|^2 \notag\\
&=& \left\| \sigma_{\max} (\mathbf A) \mathbf v_{r, \max}^H (\mathbf A) \mathbf{U} \sum_{i} \mathbf v_{l,i} (\mathbf B)
\sigma_i (\mathbf B) \mathbf v_{r,i}^H (\mathbf B) \right\|^2 \notag\\
&\stackrel{(a)}{=}& \sum_i \left\| \sigma_{\max} (\mathbf A) \mathbf v_{r, \max}^H (\mathbf A) \mathbf{U}  \mathbf v_{l,i} (\mathbf B)
\sigma_i (\mathbf B) \mathbf v_{r,i}^H (\mathbf B) \right\|^2 \notag\\
&\geq& \left\| \sigma_{\max} (\mathbf A) \mathbf v_{r, \max}^H (\mathbf A) \mathbf{U}  \mathbf v_{l,\max} (\mathbf B)
\sigma_{\max} (\mathbf B) \mathbf v_{r,\max}^H (\mathbf B) \right\|^2 \notag\\
&\stackrel{(b)}{=}& \lambda_{\max} (\mathbf A) \lambda_{\max} (\mathbf B) \left| \mathbf v_{r,\max}^H (\mathbf A) \mathbf U \mathbf v_{l, \max} (\mathbf B) \right|^2,
\end{eqnarray}
where $\sigma_i (\mathbf A)$ denotes the $i$'th singular value  of $\mathbf A$, and $\sigma_{\max} (\mathbf A)$ denotes the singular value  of $\mathbf A$ with the highest norm.  Here, $(a)$ follows from the fact that as $\left\{ \mathbf v_{r,i} (\mathbf B) \right\}$ are orthogonal vectors, the square-norm of their summation is equal to the summation of their square-norms. $(b)$ results from the fact that $\lambda_{i} (\mathbf A) = |\sigma_{i} (\mathbf A) |^2, \forall i$.
By recursively applying (\ref{mimo2}), it follows that
\begin{eqnarray}
 \lambda_{\max} (\mathbf X_{i,i}) &\geq&  \lambda_{\max} \left(\mathbf H_{K+1, \mathrm p_i(l_i -1)}\right)  \gamma_{i,l_i-1} \lambda_{\max} \left( 
 \mathbf H_{\mathrm p_i(l_i -1), \mathrm p_i(l_i -2)} \right) \gamma_{i,l_i-2} 
 \cdots \gamma_{i,1}  \lambda_{\max} \left(\mathbf H_{\mathrm p_i(1), 0} \right) \notag\\
&=& \prod_{j=1}^{l_i} \lambda_{\max} \left( 
 \mathbf H_{\mathrm p_i(j), \mathrm p_i(j-1)} \right) \prod_{j=1}^{l_i-1} \gamma_{i,j}.
\end{eqnarray}
Noting the definitions of $\mu_{\{i,j\}}$ and $\nu_{i,j}$, $(d)$ easily follows.  Finally, $(e)$ results from the fact that as $P \to \infty$, the term $\log\frac{c \left(2^{SR} -1
\right)}{P}$ can be ignored.

Since the left and the right unitary matrices resulting
from the SVD of an i.i.d. complex Gaussian matrix are independent of
its singular value matrix \cite{random_matrix_book} and $\mathbf
U_{i,j}$ is an independent isotropically distributed unitary matrix, we
conclude that all the random variables in the set $ \left\{ \left\{\mu_e\right\}_{e \in E},
\left\{\nu_{i,j}\right\}_{1 \leq i \leq L, 1 \leq j < l_i} \right\}$ are
mutually independent. From the 
probability distribution analysis of the singular values of circularly symmetric Gaussian matrices in \cite{zheng_tse}, we can easily
prove $\mathbb P \left\{\mu_e \geq \mu_e^0 \right\} \doteq P^{-N_a
N_b\mu_e^0}=P^{-w_e\mu_e^0}$. Similarly, as $\mathbf U_{i,j}$ is
isotropically distributed, it can be shown that $\mathbb P
\left\{\nu(i,j) \geq \nu_0(i,j) \right\} \doteq P^{-\nu_0(i,j)}$.
Hence, defining $\boldsymbol \mu = [\mu_e]_{e \in E}^T$,
$\boldsymbol \nu = [\nu_{i,j}]_{1 \leq i \leq L, 1 \leq j < l_i}^T$, and $\mathbf w = [w_e]_{e \in E}$,
we have
\begin{equation}
\mathbb P \left\{\boldsymbol \mu \geq \boldsymbol \mu_0, \boldsymbol \nu
\geq \boldsymbol \nu_0 \right\} \doteq P^{-\left( \mathbf 1 \cdot \boldsymbol \nu + \mathbf w \cdot \boldsymbol \mu \right)}.
\end{equation}

Let us define $\mathcal R$ as the region in $\mathbb R^{\left| E
\right| + \sum_{i=1}^{L}{l_i}-L}$ of the vectors $\left[\boldsymbol
\mu^T \boldsymbol \nu^T \right]^T$ such that for all $1\leq i \leq
L$, we have $\sum_{j=1}^{l_i} \mu_{\left\{\mathrm p_i(j), \mathrm
p_i(j-1)\right\}} + \sum_{j=1}^{l_i-1} \nu_{i,j} \geq 1$. Using the
same argument as in the proof of Theorem \ref{thm:DMT-n-ir}, we
conclude that $\mathbb P \left\{ \mathcal R \right\} = \mathbb P
\left\{ \mathcal R \bigcap \mathbb R_+^{\left| E \right| +
\sum_{i=1}^{L}{l_i}-L} \right\}$. Hence, defining $\mathcal R_+ =
\mathcal R \bigcap \mathbb R_+^{\left| E \right| +
\sum_{i=1}^{L}{l_i}-L}$ and $d_0=\displaystyle\min_{[\boldsymbol
\mu^T \boldsymbol \nu^T]^T \in \mathcal R_+} \mathbf w \cdot
\boldsymbol \mu + \mathbf 1 \cdot \boldsymbol \nu$, which can 
easily be verified to be bounded, and applying the same argument as in
the proof of Theorem \ref{thm:DMT-n-ir}, we have
\begin{eqnarray}
\mathbb P \left\{ \mathcal E \right\} \dot \leq \mathbb P \left\{ \mathcal R_+ \right\}  \doteq P^{-d_0}.
\end{eqnarray}
To complete the proof, we have to show that $d_0=d_G$, or equivalently, $d_0=L$ (note that $L=d_G$). The value of
$d_0$ is obtained from the following linear programming optimization
problem
\begin{IEEEeqnarray}{ll}
\min ~ & \mathbf w \cdot \boldsymbol \mu + \mathbf 1 \cdot \boldsymbol \nu  \\
s.t. ~ & \boldsymbol \mu \geq \mathbf 0, \boldsymbol \nu \geq \mathbf 0,
\forall i \sum_{j=1}^{l_i} \mu_{\left\{\mathrm p_i(j), \mathrm p_i(j-1))\right\}} + \sum_{j=1}^{l_i-1} \nu_{i,j} \geq 1. \nonumber
\end{IEEEeqnarray}
According to the argument of linear programming \cite{linear_book},
the solution of the above linear programming problem is equal to the
solution of the dual problem which is
\begin{IEEEeqnarray}{ll}
\max ~ & \sum_{i=1}^{L} f_i \label{eq:dual_prob}\\
s.t. ~ & \mathbf  0 \leq \mathbf f \leq \mathbf 1, \forall e \in E, \sum_{e \in \mathrm p_i} f_i \leq w_e. \nonumber
\end{IEEEeqnarray}
Let us consider the solution $\mathbf f_0 = \mathbf 1$ for
\eqref{eq:dual_prob}. As the path sequence $\left( \mathrm p_1,
\mathrm p_2, \dots ,\mathrm p_L \right)$ consists of the paths that
form the maximum flow in $\hat G$, we conclude that for every $e \in
E$, we have $\displaystyle \sum_{ e \in \mathrm p_i} 1 \leq w_e$.
Hence, $\mathbf f_0$ is a feasible solution for
\eqref{eq:dual_prob}. On the other hand, as for all feasible
solutions $\mathbf f$ we have $\mathbf f \leq \mathbf 1$, we
conclude that $\mathbf f_0$ maximizes \eqref{eq:dual_prob}. Hence,
we have
\begin{equation}
d_0 = \min ~ \mathbf w \cdot \boldsymbol \mu + \mathbf 1 \cdot \boldsymbol \nu \stackrel{(a)}{=} \max ~ \sum_{i=1}^{L} f_i = L = d_G.
\end{equation}
Here, $(a)$ results from duality of the primal and dual linear
programming problems. This completes the proof.
\end{proof}

\textit{Remark 9-} It is worth noting that according to the proof of  
Theorem 7, any RS scheme achieves the maximum diversity of the  
wireless multiple-antenna multiple-relays network as long as its  
corresponding path sequence includes the paths $\mathrm p_1,\mathrm  
p_2,\dots,\mathrm p_{d_G}$ used in the proof of Theorem 7.

Theorem 7 shows that the RS scheme is capable of exploiting the
maximum achievable diversity gain in multiple-antenna multiple-relay
wireless networks. However, as the following example shows, the RS
scheme is unable to achieve the maximum multiplexing gain in a general
multiple-antenna multiple-node wireless network.

\textit{Example-} Consider a two-hop relay network consisting of
$K=4$ relay nodes. The transmitter and the receiver are equipped
with $N_0=N_5=2$ antennas, while each of the relays has a single
receiving/transmitting antenna. There exists no direct link between
the transmitter and the receiver, i.e. $\{0,5\} \notin E$. For the
sake of simplicity, assume that the relays are non-interfering, i.e.
$1 \leq a \leq 4, 1 \leq b \leq 4, \{a,b \} \notin E$. Let us
partition the set of relays into $\mathcal S_0=\{1, 2 \}, \mathcal
S_1=\{3, 4 \}$. Consider the following amplify-and-forward strategy:
In the $i$'th time slot, the relay nodes in $\mathcal S_{i \mod 2}$
transmit what they have received in the last time slot, while the
relay nodes in $\mathcal S_{(i+1) \mod 2}$  receive the
transmitter's signal. It can be easily verified that this scheme
achieves a maximum multiplexing gain of $r=2$. However, the proposed
RS scheme achieves a maximum multiplexing gain of $r=1$.

\section{Conclusion}
The setup of a multi-antenna multiple-relay network is studied in this paper. Each
pair of nodes are assumed to be either connected through a
quasi-static Rayleigh fading channel or disconnected. A new scheme
called \textit{random sequential} (RS), based on the
amplify-and-forward relaying, is introduced for this setup. It is
proved that for the general multiple-antenna multiple-relay
networks, the proposed scheme achieves the maximum diversity gain.
Furthermore, bounds on the diversity-multiplexing tradeoff (DMT) of
the RS scheme are derived for a general single-antenna
multiple-relay network. Specifically, 1) the exact DMT of the RS scheme
is derived under the assumption of ``non-interfering relaying''; 2)
a lower-bound is derived on the DMT of the RS scheme (no conditions
imposed). Finally, it is shown that for the single-antenna
two-hop multiple-access multiple-relay network setup where there is no
direct link between the transmitter(s) and the receiver, the RS
scheme achieves the optimum diversity-multiplexing tradeoff.
However, for the multiple access single relay scenario, we show that the RS scheme is unable to perform optimum in terms of the DMT,
while the dynamic decode-and-forward scheme is shown to achieves the optimum DMT for
this scenario.

\bibliographystyle{IEEEtran}
\bibliography{relay_tradeoff_submitted}
\end{document}